# Title

Climate change impacts on supra-permafrost soil and aquifer hydrology: broader, deeper, and longer activity

# Authors

Neelarun Mukherjee[1*], Bo Gao[2], Ethan T. Coon[2], Pin Shuai[3], Devon Hill[3], Bethany T. Neilson[3], George W. Kling[5], Jingyi Chen[1,6], and M. Bayani Cardenas[1]

# Affiliations

[1]Department of Earth and Planetary Sciences, The University of Texas at Austin, Austin, TX

[2]Environmental Sciences Division, Oak Ridge National Laboratory, Oak Ridge, TN

[3]Civil and Environmental Engineering, Utah Water Research Laboratory, Utah State University, Logan, UT

[4]Department of Earth and Environmental Sciences, University of Michigan, Ann Arbor, MI

[5]Department of Ecology and Evolutionary Biology, University of Michigan, Ann Arbor, MI

[6]Department of Aerospace Engineering and Engineering Mechanics, The University of Texas at Austin, Austin, TX

*correspondence: neelarun@utexas.edu

# Abstract

The thermal dynamics and hydrology of active layer soils and supra-permafrost aquifers determine the fate of the vast pool of carbon that they hold. In permafrost watersheds of Arctic Alaska, air temperature has warmed by up to 3.5 °C and snowfall has increased by up to ~40 mm from 1981 to 2020. How these changes impact the seasonal to decadal hydrological activity of the carbon-rich aquifers is mostly unknown. Observation-informed thermal hydrology modeling of a hillslope drained by a headwater stream (Imnavait Creek) within continuous permafrost showed profound changes from 1981 to 2020. Warmer summer temperatures deepened annual thaw depths. Steadily warming winter air temperatures, heavier snowfall, and stored energy from summer increased annual water outflow from the hillslope aquifer to the stream, warmed soil temperatures, and expanded and prolonged zero-curtain (stable at 0 °C) zones. In 2017-2018, zero-curtain areas with liquid water persisted through winter. Our findings reveal that both summer and winter warming drive year-round aquifer dynamics, creating conditions that amplify the permafrost-carbon-climate feedback.

# Teaser

Arctic winter warming and warmer summer temperatures have crossed a tipping point from frozen to thawed and expanded and sustained year-long hydrological activity above permafrost.

**MAIN TEXT**

**INTRODUCTION**

Permafrost occupies around 25% of the northern hemisphere and 80% of arctic watersheds (*1, 2*), and its soils store twice as much carbon (C) as currently present in the atmosphere (*3, 4*). The Arctic is warming almost four times the rate of the rest of the world (*5–7*), particularly during winter (*8–12*). This rapid warming can potentially accelerate both the spatial and temporal extent of thawed permafrost, exposing previously frozen organic C to hydrological, chemical, and biological processes. As permafrost thaws, soil particulate organic matter leaches into the groundwater within shallow aquifers as dissolved organic carbon (DOC). Microbial decomposition of DOC releases $CO_2$ and $CH_4$ to the atmosphere, further increasing warming (*6, 13, 14*) by activating a strong, positive permafrost-carbon-climate feedback (*4, 6, 15–18*). However, despite the recognized importance of these feedbacks, the physical mechanisms controlling how the long-term climatic change alters the subsurface thermal and hydrological regimes remain poorly characterized.

Feedbacks between climate and soil carbon dynamics are fundamentally governed by the subsurface thermal and hydrological environment, which is in turn strongly influenced by long-term climate change. Long-term meteorological data (*19*) across multiple Arctic Alaska sites – watershed of Imnavait Creek, the Kuparuk River, the Colville River, and the Sagavanirktok River - revealed a consistent and significant warming trend over the past forty years. All sites exhibited strong winter (September - May) amplification of warming. The long-term temperature data highlighted that the coldest years were clustered in the earlier part of the record, while the warmest winters consistently appeared after 2005, reflecting a persistent and directional climatic shift. Along with warming trends over the winter, long-term records showed a positive trend in both cumulative winter snowfall over the 1981-2020 period (Fig. 1E). These trends have major implications for freeze-thaw cycles and the dynamics of subsurface hydrology in Arctic Alaska.

Despite clear evidence of warming and increasing snowfall patterns, we understand only the basics of the impact of climate change on subsurface hydrological and thermal states, critical drivers of C fluxes (e.g., *14, 15, 20–26*). During summer, the active layer thaws and becomes a shallow, unconfined aquifer (a supra-permafrost aquifer). The magnitude of the fluxes of C from land to surface waters and to the atmosphere is strongly governed by the hydrologic and thermal state of this aquifer (*23, 27*). Deeper thaw under warmer summers enhances microbial activity, while frequently saturated active layers increase hydrologic connectivity, flushing more DOC into streams or emitting more C gases to the atmosphere (*28, 29*). In winter, even if the surface of the soil freezes, unfrozen zones at depth can still persist within the soil where temperature is stable near 0 ºC (zero-curtain, ZC, zones) due to latent heat exchange and to insulation from snow cover (*30, 31*). These zones can sustain slow but steady microbial activity (*32, 33*) due to warmer and wetter conditions. Observations from arctic tundra ecosystems have shown that C fluxes can be high during the early summer, likely due to delayed release of stored gas in the ZC zone from the previous season's summer (*34*), and cold-season gas emissions to the atmosphere may rival or even exceed the summer growing season $CO_2$ uptake by land (*13, 32, 35–38*). However, most of these studies have largely inferred belowground mechanisms from surface gas flux observations, and we lack the mechanistic knowledge on how climatic shifts control subsurface thermal and hydrological states, how year-round C transformations and fluxes are sustained, and how aquifer dynamics will alter soil C cycling in the future.

Bridging these knowledge gaps requires a mechanistic understanding of how permafrost thaw alters groundwater storage and flow within supra-permafrost aquifers. Field and modeling studies in arctic watersheds have long noted that a significant fraction of streamflow during thaw seasons is sustained by groundwater from the aquifer (*39*). Techniques such as hydrograph separation (e.g., *39, 40*) and the use of geochemical tracers (e.g., *41*) have been used to estimate groundwater contributions to streams. However, field studies are limited by sparse temporal and spatial coverage; it is especially difficult to make high-resolution observations in the subsurface during winter. Process-based models provide an alternative for investigating thermal-hydrological dynamics in soils experiencing temporal freeze-thaw. Process-based models such as the Advanced Terrestrial Simulator (ATS, *42*) can capture coupled permafrost thermal-hydrological dynamics, but they require detailed calibration in order to replicate field situations (*43*, *44*). Previous arctic studies using multi-physics models showed that warming increases the groundwater contribution to streamflow (*45*), that lateral groundwater heat advection and topography alter freeze-thaw dynamics (*46–48*), and that snow cover and air temperature strongly affect the aquifer's thermal state (*49*). Another modeling study demonstrated the influence of continentality, temperature, and precipitation on permafrost degradation (*50*). These computational studies illustrate the sensitivity of these aquifers to a changing climate, yet few of these virtual experiments have been validated by co-located field observations. Prior studies and our preliminary work show that thermal hydrology models of soils and aquifers, especially the relatively thin and shallow ones in the Arctic, struggle to faithfully represent observations (*44*, *51*, *52*). These discrepancies are not mainly due to inadequate model physics, but are driven by biases, errors, and inconsistencies in the hydro-meteorological inputs into models. Recent, direct field observations on the spatial distribution of soil layer thicknesses, hydraulic properties, and thermal properties at the landscape scale (*53*, *54*) can help constrain thermal hydrology models. Computational modeling experiments using coupled thermal-hydrological models that are manually calibrated against co-temporal measurements are needed to understand the governing physical, thermal, and hydrological processes in response to climate change.

To address this need for field-data calibrated models to determine mechanisms of aquifer behavior, we show how long-term climate change affects the supra-permafrost aquifer dynamics in a continuous permafrost headwater catchment. We collected high-resolution, co-located thermal and hydrologic observations over seasonal cycles. These data were synthesized with previous measurements of thermal and hydrologic properties to inform a high-resolution ATS multiphysics model. The model was applied for a 2D vertical transect, spanning a hillslope to a riparian zone, to determine how seasonal and interannual meteorology and climate control groundwater flow, subsurface water saturations, and freeze-thaw cycles in a representative aquifer (Fig. 2B). The study site, Imnavait Creek, is a typical headwater watershed in continuous permafrost on the North Slope of Alaska. An iterative model-data integration framework ("ModEx"; *51*) using both historical (*53*, *55–58*) and high-frequency (spatial and temporal) field observations constrained the model physics, parameters, and modeling structure, and the model itself was able to reproduce field observations. The ATS modeling was implemented with daily forcing hydroclimate data during 1981-2020. The historical hydroclimate data was analyzed to detect changes over the 40-year period.

**RESULTS**

Four decades of persistent warming, and increased snowfall at the Imnavait Creek watershed (Fig. 1) have changed the thermal and hydrologic regimes of the supra-permafrost aquifer, especially during winter. Imnavait Creek showed a winter temperature increase of +2.98 °C (Fig. 1, A and C) and its broader watershed, the Kuparuk River, showed a +2.76 °C increase (Fig. 1, A and C). Neighboring major watersheds also warmed up; by +3.47 °C at the Colville (Fig. S1, A and D), and +2.89 °C at the Sagavanirktok (Fig. S2, A and D) watersheds. Summer (June to August) temperatures showed a moderate increase: 0.42 °C at Imnavait Creek (Fig. 1, A and D), 0.59 °C at the Kuparuk watershed (Fig. 1, A and D), 0.50 °C at Colville (Fig. S1, A and E), and 0.53 °C at Sagavanirktok (Fig. S2, A and E).

By combining long-term climate forcing and site-calibrated process modeling, this work links decadal climate trends directly to thermal and hydrological processes. The model reproduced key aquifer hydro-thermal dynamics in terms of groundwater flow, freeze-thaw cycles, snowpack insulation, as well as the formation of ZC zones, regions within the soil where temperatures remain close to 0 °C despite declining and sub-freezing surface and atmospheric temperatures. This model reproduction of data enabled us to evaluate critical thermal and hydrological responses in supra-permafrost aquifers across both intra- and inter-seasonal timescales.

**Intra-year Thermal and Hydrological Responses**

*Winter* - The simulations showed how the ground lost heat to the atmosphere with decreasing winter air temperatures. However, the snowpack acted as a thermal insulator, substantially reducing subsurface heat loss. This insulation limited ground cooling and helped maintain temperatures within the aquifer close to 0 °C throughout some winters. Field temperature data from vertical thermistor arrays installed at four locations (labeled as IMP-05, IMP-09, IMP-14, and IMP-18 in Fig. S3) confirmed that the aquifer remained near 0 °C for much of the winter (Fig. S3). These observations were also reproduced in the model, which showed shallow soil temperatures typically ranging from 1 °C to -4 °C depending on location and depth (Fig. 3 and Mov. S1). Some downslope locations along the transect remained above the freezing point with a higher liquid water content throughout winter (Mov. S1). Overall, as freezing progressed bidirectionally from both the land surface above and the permafrost table below (Mov. S1), sandwiched unfrozen zones near 0 °C (or ZC; *30*) were temporarily present over a sustained period of time. This period is the zero-curtain period (or ZCP).

Despite the presence of ZC zones, the subsurface remained mostly fully frozen, and winter hydrological activity was minimal. The frozen surface made infiltration effectively zero. Winter precipitation (snowfall) accumulated as snowpack above the frozen ground. No active vertical or lateral groundwater flow occurred within the frozen soil. Hydrological fluxes across the land surface were negligible across the transect during these months, confirming hydrological dormancy (Fig. 4A and Mov. S1). Within the ZC zones, no substantial hydrologic connectivity occurred through the winter months, i.e., there were no continuous flowpaths from hillslope to the stream.

*Start of Thaw and Early Summer* - During the months of May to June, rising air temperatures and increasing solar radiation initiated a rapid transition in thermal and hydrological regimes. Thawing began at the surface and progressively deepened, but in the early part of the season the subsurface remained largely frozen below the topmost peat layer, or the acrotelm layer (Fig. 5). The onset of snowmelt and the resulting freshet typically occurred in May when air temperatures rose above 0 °C, which led to rapid ablation of the seasonal snowpack (Fig. 4, A and B and Mov. S2). Thermal profiles during this period showed rapid warming in the upper 10 to 20 cm of the soil column, with temperature gradients continuously steepening as the thaw front deepened (Fig. 5). Simultaneously, the snowpack continued to thin and lose its insulating effect (Mov. S2), and the thawed aquifer temperature increased with the deepening of thaw. The temperature of the subsurface below the ice table remained negative but very close to 0 °C (Fig. 5).

Infiltration capacity at this time is limited because the aquifer is very thin (one to several cm in the upper vegetation layers). Most of the meltwater flowed downslope as saturation-excess overland flow and resulted in a pronounced outflow to the stream (Fig. 4B and Mov. S2). Despite shallow thaw depths, the highest subsurface flow rates were also observed during early summer (Figs. 4C, 5, and Mov. S2). This was attributed to the thawed surficial acrotelm, characterized by high hydraulic conductivity that facilitated rapid lateral flow down the topographic gradient and surface-subsurface exchange. In some instances, the freshet coincided with early-season rainfall events, producing sharp infiltration pulses that rapidly increase shallow aquifer storage (Mov. S2). These pulses were followed by quick drainage and then by progressive drying from upslope to downslope locations as the storm event subsided (Mov. S2). With an increase in thaw from early-summer to mid-summer, infiltration increased, and the aquifer began to drain more steadily. Subsequent storm events during this period led to near-instantaneous saturation of the shallow aquifer, followed by saturation-excess overland flow following the surface topographic gradient with less substantial surface-subsurface exchange (Mov. S2). However, following prolonged dry periods the model indicated that the initial response to a new storm is increased infiltration. This is evidenced by downwards (negative) surface boundary normal fluxes (Fig. 4C). This occurs before there is sufficient saturation needed to re-establish large lateral flows (Mov. S2). These intra-seasonal dynamics highlighted the interactions between thaw progression, snowmelt intensity, and antecedent moisture conditions in controlling the hydrologic response of the aquifer.

*Late Summer* - During the months of July to September, the aquifer had maximum thermal development, with progressive warming and deepening of the thaw front (Fig. 5). Surface soil temperatures exceeded 10 °C in July and August following air temperatures, while temperatures at the base of the aquifer in mineral soil remained near 0 °C due to latent heat effects and proximity to underlying permafrost (Fig. 6). The acrotelm and deeper catotelm (older peat) warmed rapidly due to their low heat capacity, while the mineral soil layers underlying the peat layers responded more gradually (Fig. 5). This created a steep vertical temperature gradient with downward conductive heat flux. By the end of August, the temperature gradient began to flatten slightly, indicating a transition toward thermal equilibrium within the aquifer before the onset of early winter cooling (Figs. 5 and 6). These model results were consistent with field observations (Fig. 3) and showed the characteristic summer thermal regime of Arctic permafrost hillslopes: a warm, fully-thawed upper layer, underlain by persistent near-freezing temperatures due to colder, deeper permafrost, latent heat buffering, and both thermal heat transfer and insulation from overlying organic soils (Fig. 6).

The subsurface hydrologic system became fully connected during the summer period. The infiltration from rainfall was the primary water input. Summer was characterized by dynamic hydrologic responses to rainfall events and strong surface-subsurface coupling, evident from higher positive and negative values of surface normal flux laterally along the 2D transect (Fig. 4D and Mov. S2). By late July, all the layers of the supra-permafrost aquifer - the acrotelm, catotelm, and the top mineral soil (where present) - were thawed. Vertical profiles of magnitudes of volumetric groundwater flux per unit area (sometimes referred to as the Darcy velocity, $|\vec{V_l}|$ (Eq. 3 in Supplementary Materials)) showed sharp differences between the three layers because of their different hydraulic properties (Fig. 5). Groundwater flux magnitudes in the acrotelm were almost two orders of magnitude higher than groundwater fluxes in the catotelm, which was again one to two orders of magnitude higher than that in the mineral soil (Fig. 5 and Mov. S2). During large storms, the aquifer became fully saturated, which led to rapid lateral flow with overland runoff (Fig. 4D and Mov. S2). Groundwater fluxes during these events showed a strong downslope-directed tangential component (with flux magnitudes reaching ~1 m/d), indicating efficient subsurface transport through the highly permeable acrotelm and catotelm (Mov. S2). The model also showed the sensitivity of flow pathways to antecedent moisture conditions. After extended dry periods, the initial phase of a storm was dominated by infiltration, as evidenced by strong negative vertical fluxes at the surface (Fig. 4D and Mov. S2). Once the shallow soils were saturated, vertical fluxes from the subsurface to surface boundary occurred (Fig. 4D). Most of the flow is downslope but there are some local surface-subsurface exchanges (Fig. 4D and Mov. S2). These transitions were particularly evident in profiles of magnitudes and pore-space liquid water saturation at x = 105 m (distance from the stream), where groundwater fluxes increased by several orders of magnitude following intense rainfall (Mov. S2). Preferential flowpaths emerged along topographic and stratigraphic gradients, with localized zones of convergence and divergence evident in the groundwater flux field mostly during a dry period after a storm event (Fig. 4 and Mov. S2). These patterns suggest that both surface topography and underlying soil texture cause localized low-hydraulic head areas influencing flow distribution. For example, convergence zones at downslope positions corresponded to areas of increased saturation and ponding of surface water (Mov. S2), consistent with field observations as well as with previous studies of fully-saturated riparian zones in the area (*24*).

*Freeze-up* - As the summer progressed into autumn and early winter (September - October), the aquifer cooled in response to snow-precipitation, declining solar radiation, and subzero air temperatures. Temperature profiles from September and October showed the onset of surface cooling; lower air temperatures cooled the soil surface (Fig. 6). Despite this surface cooling, most of the aquifer remained near 0 °C due to latent heat release from refreezing (Fig. 5 and 6). This initiated a bidirectional freeze-up where freezing propagated from both the surface downward and the permafrost table upward (Figs. 4E, 5, 6 and Mov. S1). This caused the formation of the ZC zone, where latent heat release associated with the phase change of water to ice buffers the local thermal gradient and delays further cooling.

The timing and persistence of the ZC varied spatially along the transect due to lateral differences in soil moisture content (Fig. 6). At upslope locations (IMP-01 to 05 in Fig. 2), where soils were relatively dry following summer drainage, the aquifer froze more rapidly (Fig. 6). The smaller amount of liquid water filling the pores reduced latent heat effects due to lower effective thermal capacity, allowing for a quicker drop in temperature and early onset of complete freezing. In contrast, downslope riparian zones where soils retained more water had prolonged ZC conditions. The greater water availability led to extended phase-change buffering, slowing the progression of freezing despite similar air temperature exposure. As a result, the ZC emerged earlier and was shorter-lived in drier upslope soils, while in wetter downslope soils it formed later but persisted longer into winter (Fig. 6 and Mov. S1). The model showed this spatial variability in freeze-up timing and thermal buffering, demonstrating that soil moisture distribution strongly influenced the thermal evolution of the aquifer during autumn and led to asynchronous freezing patterns across the hillslope-riparian continuum (Fig. 6 and Mov. S1).

Groundwater flow in the aquifer underwent a gradual shutdown in connectivity during the bi-directional freeze up of the soil (Fig. 5, 6 and Mov. S1). This bidirectional freezing increasingly isolated the remaining unfrozen zones, limiting both vertical and lateral flow. The period during late October showed a marked reduction in groundwater fluxes throughout the domain, with tangential (downslope) and normal (vertical) components dropping below 1 cm/d (Fig. 5). The freeze-induced decline in hydraulic conductivity, driven by ice formation in pore spaces, greatly restricted water movement particularly in upslope locations where soil water content was already low. At upslope and midslope areas (IMP-01 to 07 in Fig. 2), the aquifer was relatively dry by late summer due to efficient drainage and lower water retention capacity. As a result, freezing proceeded rapidly and these regions transitioned to a hydrologically disconnected state early in autumn. Minimal infiltration occurred during early summer storms, and any precipitation was either shed as overland flow or retained near the surface and rapidly frozen. Downslope riparian zones maintained higher antecedent moisture due to convergent flow and topographic accumulation and stayed thawed longer. These wetter soils provided higher effective heat capacity and latent heat buffering, delaying complete freeze-up and enabling some lateral flow even at the start of freezing. With more freezing, the lateral fluxes decreased due to closure of pore spaces by ice (Fig. 5).

**Interannual Trends in Thermal and Hydrological Response (1981-2020)**

*Interannual Variability in Subsurface Thermal Profiles During Winter* - The annual duration of near-zero soil temperatures (±0.2 °C around 0 °C; ZC period or ZCP, Fig. 7A) increased substantially over the 40-year simulation. To assess long-term trends in mid-winter ground thermal conditions, we extracted both the ZCP (Fig. 8A) and transect temperature on January 22 at each location for each year from 1981 to 2020 (Fig. 8B). January 22 is the average climatological coldest day of the year across the 40-year simulation period.

During the earliest years (1981-1990), the upslope and midslope regions had a shorter ZCP of ~100 days, and the downslope areas had a longer ZCP on average, ranging from 170 days close to the creek and 90-100 days near the midslope regions (Fig. 8A). During winter, land surface temperatures followed air temperatures, and below-ground temperatures ranged from -4 °C to -8 °C (Figs. 6 and 8B). The unfrozen ZC zone remained only near the riparian zone with a small spatial extent (~1 to 5 m from the stream at the downslope end, and 3 to 5 cm vertically, Figs. 6 and 8B).

Starting in the mid-1990s and continuing through the 2000s, there was a shift toward warmer temperatures and longer ZCP. For example, the 1993-1995 winters saw consecutive ZCP of more than 100 days throughout the transect (Fig. 8A). Surface temperatures on January 22$^{nd}$ began to exceed -5 °C, and the temperatures at 20-60 cm depth were progressively warmer, typically ranging between -2.5 °C and -4 °C (Fig. 8B). This was accompanied by a flattening of the vertical thermal gradient, suggesting reduced conductive cooling during winter (Fig. 5). The years from 1993 to 1996 showed a relatively warmer subsurface, with a thicker ZC zone compared to the previous decade, spanning up to 6 to 8 cm vertically and ~5 to 10 m horizontally from the downslope end (Fig. 8B). On January 22$^{nd}$ in 1994 and 1995 the entire frozen aquifer had a temperature range from -2 °C to -0.1 °C (Fig. 8B). These observations were also consistent with warmer average winter temperatures from 1993 to 1996 (Fig. 1C).

The decade from 2001 to 2010 also showed a similar trend with some warmer winter years (2004 to 2006, Fig. 1C), and winter aquifer temperatures lingering close to 0 °C (Fig. 8B). This decade showed a somewhat cooling pattern compared to 1990 - 2000. This pattern was also correlated with the average yearly winter air temperatures as well as ZCP along the transect, both showing a decreasing trend from 2000 to 2009. Despite being colder than average winters, the 2002-2005 winters displayed longer ZCPs, lasting >100 days in upslope and >170 days in downslope zones (Fig. 8A), due to stored thermal energy (enthalpy) from warmer summers (Figs. 7F and S5, see below). The 2007-2008 winter season was the coldest of the decade, but it had less than 100 days of ZCP throughout the transect except for a very small location in downslope area (Fig. 8A).

After 2010, ZCP durations lengthened markedly, with many winters exceeding 100 ZCP days (Fig. 8A). Prior temperature observations at four locations in the riparian zone from 2013-2014 (Fig. S4) corroborate the modeled temperatures and ZCP; the modeled ZCPs (Fig. 8A) bound the observed ZCPs for 2013-2014, which ranged from 183 to 288 days (Fig. S4). By the late 2010s, near-continuous unfrozen conditions in mid-soil depths persisted through most of the cold season. This trend culminates in extreme years such as the 2017-2018 winter, when the transect maintained extensive ZC conditions (>180 days) across its full length (Fig. 8 and Mov. S1). The warming trend steepened during this time period. Several temperature profiles showed surface temperatures above -2 ºC and mid-depth values approaching -0.5 °C to -2 °C. In some years, the upper 50 cm of the soil profile became near isothermal around 0 °C (Fig. 8B). The ZC zone also expanded by ~20 m horizontally from the downslope end of the transect and by 5-15 cm vertically (Fig. 8B). The warmer winter of 2018 showed the presence of a continuous unfrozen layer across the entire hillslope-riparian zone model domain (Fig. 8B). The winter of 2017-2018 was one of the warmest winters on record (Fig. 1C), which prevented the supra-permafrost aquifer from freezing completely (Fig. 8, B and C). The modeled and measured observations shift systematically towards warmer cold-season soil conditions, likely due to the combined effects of atmospheric warming, thicker or more persistent snowpacks, and accumulated thermal energy from warmer preceding seasons.

Summer maximum enthalpy (subsurface heat content) increased from 1985 to 2020, with a persistent and relatively steep increase after 2008 (Fig. 7F). The increase in maximum summer enthalpy is correlated with an increase in the spatial extent and duration of the zero-curtain during the following winter season (Fig. 7F). These relationships demonstrate thermal memory, where enhanced freeze–thaw buffering in the winter accompanies greater energy accumulation in the soil in the preceding late summer.

*Deepening of the aquifers in Warm Summers* - Simulated thaw depths showed that summers with anomalously high mean air temperatures cause more thawing and produce a deeper aquifer along the hillslope-to-creek transect (Fig. 9). While the long-term summer warming trend was modest (~ 0.4 to 0.6 °C over four decades, Fig. 1), some warmer summers such as 1990, 2004, 2007, 2010, 2017, and 2019 produced thaw depths that exceeded the forty-year mean thaw depth (68.2 cm from the model) by more than 10 cm. Years 1990, 2004, 2007, 2010, 2017 and 2019 had maximum thaw depths of 78, 84, 86, 79, 77, and 80 cm, respectively (Fig. 9). On the other hand, the shallowest aquifer occurred in 2002 (51 cm), 2000 (53 cm), and 1981 (55 cm), coinciding directly with cooler summers with mean summer temperatures below 6 °C. The overall interannual range of thaw depths was large, from 51 to 86 cm, reflecting strong interannual variability in summer temperatures. A strong positive correlation exists between mean summer temperature and maximum aquifer thickness, with thickness increasing by 6.59 cm per °C increase in mean summer temperature (Fig. 9). The four-decade summer warming at our field site led to an increase in maximum thaw depths by 0.23 cm per year (Fig. 9).

*Relationships Between ZCP, Winter Temperatures, and Hydrologic Response* - Across the 40-year period from 1981 to 2020, the maximum unfrozen area increased over time (Fig. 8C). We analyzed the mean ZCP over all model cells and the mean ZC area for that period throughout the study domain (Fig. 7C). In earlier years (1981 -1995), the mean subsurface ZC covered 37 % to 42 % of the maximum unfrozen area and lasted 60-90 days per year (Fig. 7, B and C). This period included both early summer thaw and early winter freeze-up intervals. However, starting around 2009, the ZC showed more areal expansion (50 to 55 % of maximum unfrozen area) and the ZCP lengthened markedly (~100 days, Fig. 7, B and C). In the final decade (2010 - 2020), the ZCP frequently exceeded 120 to 130 days (Figs. 7B and 8A) with consistently more than 50 % of the aquifer unfrozen (Fig. 7, B and C and 8, A and B). The spatiotemporal trend in ZCP showed a clear and steady increase over time, consistent with long-term atmospheric warming. A positive correlation was observed between the ZCP length and area with average winter air temperatures as well as cumulative snow precipitation across years (7B). On the other hand, mean ZC area showed a stronger positive trend with yearly snowfall than with winter air temperature (Fig. 7C).

The timing of early summer thaw shifted earlier over the course of the simulation. In the 1980s, thaw onset typically occurred around day-of-year (DOY) 150 (late May) (Fig. 7D). In contrast, during the last decade (2010-2020) thaw initiated as early as DOY 120 (late April). This earlier shift in thaw timing was inversely correlated with the mean winter air temperature (Fig. 7D). However, snow precipitation showed a positive trend with the timing of thaw. This means that as winter snowfall increased, the following summer's thaw started later. Longer periods of snowpack cover prevented an early thaw by providing insulation from the atmosphere and shading from solar radiation.

Annual cumulative outflow from the hillslope-riparian zone to the creek was also linked to mid-winter thermal conditions. Years with higher average winter temperatures and snow precipitation consistently exhibited greater total annual outflow. This pattern became more prominent in the later years of the simulation. During this period, increased outflow volumes coincided with both warmer winter air temperatures and prolonged ZCP (Fig. 7E).

**DISCUSSION**

Persistently warming winters and warm summer temperatures expand the thermal and hydrological activity of supra-permafrost aquifers both spatially and temporally, with important consequences for the permafrost-carbon-feedback. Prolonged winter-time liquid water retention potentially increases the time window for respiration and allows for some lateral C transport. Warmer summers deepen the aquifers creating more area for biogeochemical activity and DOC transport. These findings demonstrate that short term seasonal warming as well as long term inter-year warming act as dual regulators of year-round subsurface thermal and hydrologic connectivity and C-cycling in Arctic headwaters.

**Seasonal Thermal and Hydrological Dynamics and their Biogeochemical Implications**

The seasonal evolution of heat and hydrology within the aquifer reflects a balance between snowpack accumulation, soil moisture variability, and subsurface thermal gradients. Our simulations reproduced key features of this cycle and show how hydrological activation, surface-subsurface connectivity, and the persistence of unfrozen zones together control the timing and magnitude of groundwater flow. During early summer, thaw initiation and the highly permeable and porous acrotelm allows instantaneous infiltration. Advective heat transfer, aided by increased solar radiation and warmer air temperatures, warms near-surface soils and steepens vertical temperature gradients and increases thaw depths (Fig. 5). This hydrological activation of the aquifer during early summer after the frozen winter can reactivate microbial respiration in the surface acrotelm, and open pore spaces for the delayed release of stored gas in the ZC zone from the previous fall. Our simulations showed that shallow lateral flow during early summer connects unfrozen zones, and can also reconnect previous ZC zones to the surface, potentially leading to and explaining observed early season $CO_2$ and $CH_4$ release (*35, 36, 59*).

By mid to late summer, the fully thawed aquifer supports increased downslope groundwater flux and enhanced surface-subsurface exchange. Increased rain increases pronounced downslope flow and also some increased surface-subsurface exchanges driven by head differences induced by surface topography (Fig. 4). Similar observations could be explained as a small-scale version of topography-driven Tóthian regional groundwater flow (*60, 61*). However, strong contrasts in hydraulic conductivity - orders of magnitude higher in the acrotelm than in deeper soils (*62*) - limit the magnitude of exchange between the surface organic layers and the deeper mineral layers. The seasonal changes in aquifer thermal hydrology have important implications for biogeochemical cycling in supra-permafrost aquifers. Below the ice table, the temperatures were below freezing but within several degrees of 0 $^{\circ}$C (Figs. 5 and 6). This suggests an increase in pore space liquid water content just below the ice table, consistent with minimal flux of unfrozen residual water films under subzero conditions (*63*). Together, these processes highlight how summer thaw couples hydrological activation with rapid heat transfer, causing the sensitivity of Arctic subsurface systems to small shifts in seasonal energy balance.

As air temperatures declined during early winter, ZC zones with higher liquid water content appeared because of bidirectional freezing, snowpack insulation, and latent heat buffering (Fig. 5, 6 and Mov. S1). Microbial processes could potentially persist in this ZC zone during winter, as higher winter C-emissions were reported by some studies (*13, 35*).

**Impacts of Long-Term Winter Warming**

The observed record of four decades of winter warming within Arctic Alaska has led to a pronounced and sustained thermal regime shift in soils. Winter warming increased the duration as well as the extent of the ZC, particularly in downslope riparian regions (Figs. 6 and 8) where soils retain more water. These persistent near-isothermal subsurface conditions are thermodynamically significant; they indicate prolonged latent heat exchange due to phase change and result in buffering of further cooling even during peak winter. The rise in late-summer subsurface enthalpy amplifies this shift by more stored energy before freeze-up, further delaying refreezing and prolonging ZC conditions. The prolonged ZCP prevents loss of heat from the soil during winter, allowing energy to persist into the following summer (Fig. S5). This shows that the aquifer has long-term thermal memory, as opposed to seasonal temperature signals that decay rapidly. This substantial thermal memory means that winter freeze-up does not fully reset its energy state (Fig. S5), and it may contribute to persistent enhancement of gas fluxes from the tundra following years with particularly warm soils (e.g., ref 38).

The spatial expansion and lengthening of the ZCP with winters that are increasingly warming have major implications for biogeochemical cycling. The model results indicate that the lengthening of the ZCP also correlated directly with early thaw onset and more downslope outflow. Projections suggest huge losses of soil C from the top 30-50 cm of Arctic soils in the next 8 decades, most of which is expected to occur during the ZCP months from October to April (*13*). An increase in ZCP in downslope riparian areas would delay complete freeze-up and preserve liquid water in C-rich ZC zones, providing an unfrozen environment for microbial activity into winter. Our model indicates that longer ZCPs expose unfrozen C-rich zones for longer durations.

**Impacts of Deeper Thaw in Warm Summers**

Summers with high temperatures deepen the aquifer (Fig. 9), which also expands the thermally and hydrologically active volume of the subsurface. A thicker aquifer establishes new connections between organic-rich acrotelm and catotelm with the deeper mineral soil, altering water residence times and solute pathways. This can release buried, immobile carbon into hydrologically connected pathways through organic layers. While this decreases the effective residence times for chemical or biological processing, it can increase the export of deeper, older C. This older C has been shown to be microbially and photochemically labile once it reaches surface waters (*64, 65*). Similar patterns of thaw-driven changes in hydrologic pathways have been noted in arctic watersheds where thaw enhances subsurface flow and increases lateral export of deeper solutes and nutrients (*41, 66, 67*).

Deeper thaw years can potentially mobilize soil organic matter that is otherwise isolated and frozen in colder summers. The exposure of deeper and older C, along with microbes, enhances microbial decomposition and contributes to more C release. This is consistent with long-term field observations and global syntheses showing that deeper thaw accelerates the permafrost-carbon-feedback (*4, 9, 68, 69*). Vegetation productivity is tightly constrained in continuous permafrost tundra of the Arctic as well. This newly available C and its loss from soils during deeper thawing years is unlikely to be offset by vegetation uptake, potentially resulting in a net release to the atmosphere or to surface waters (*27, 70*). A deeper aquifer would have deeper flow paths, which can alter soil water chemistry and have implications for downstream aquatic ecosystems and biogeochemical cycling (*25, 41*).

These findings show that short term climate warmings are potentially as critical as the mean warming trends for aquifer dynamics. While long-term warming gradually thickens the soil active layer, individual warm years strongly impact the soil thermal regime (Fig. 7, 8) and create conditions for greater microbial activity and C export (Fig. 7). This observation is consistent with ecosystem studies showing that warmer winter soil temperatures strongly impact terrestrial C balance (*38, 71*), and that short-term weather variability can strongly constrain land-atmosphere C exchange (*18*). The following two thermal-hydrologic processes: (1) an increase in summer warmer temperatures that deepen the aquifer, and (2) an increase in winter temperatures and stored summer heat leading to greater duration and extent of unfrozen or near $0\ °C$ soil conditions, are important controls on the biogeochemistry of the permafrost-carbon-feedback.

**MATERIALS AND METHODS**

To address the question of how four decades of climate change affected the thermal and hydrologic dynamics of active layer soils, we used historical data to constrain model physics and parameters and performed manual iterative co-temporal model calibration and validated with co-located field observations from the study site.

**Study Site**

Imnavait Creek is located ~10 km east of the Toolik Field Station in the northern foothills of the Brooks Range, Alaska. It is a first-order beaded stream that drains into the Upper Kuparuk River (Fig. 2B). The 2.2 km$^2$ area of the upper Imnavait Creek watershed is among the most studied watersheds in the Arctic (for example, *23*, *24*, *38*, *52*, *72–77*). The watershed consists of relatively gentle slopes from hilltops with heath tundra down to valley bottoms with wet-sedge tundra. The hillslopes are primarily covered with tussock tundra and are intersected by "water tracks". Water tracks that begin just below the ridges are oriented parallel to the slope and feed directly into a low-gradient riparian zone and function to shorten hydrologic response time to precipitation events (*72*). During heavy precipitation events, saturation-excess overland flow is generated within water tracks that feed surface water to the main channel from the hillslopes.

Topography is the primary control on groundwater flow within the Imnavait Creek watershed, as demonstrated by previous field and modeling studies showing that subsurface water and heat fluxes in these hillslopes are largely slope-parallel and topography driven (*23*, *54*). We represent the system as a 2D vertical hillslope-riparian transect following one of these topographic-gradient aligned flow-pathways. This approach also aligns with the spatial structure of our field measurements and enabled collocated high-resolution model tuning and validation. The transect also captures the hillslope-riparian continuum typical of the North Slope watersheds, within a wet sedge, tussock tundra, and tussock-shrub tundra ecosystem (*53*, *78*). These three landcover types represent approximately 45% of the entire North Slope (Table 1 of *78*), making this transect regionally representative.

Throughout the watershed, a similar soil stratigraphy is observed in the supra-permafrost aquifer that has three distinct soil layers: surface acrotelm, mid-depth catotelm, and deeper mineral soil (*62*) (Fig. 2C). The observed thaw depth in the study region is representative of tundra located in the northern foothills of the Brooks Range, Alaskan North Slope, where the average end-of-season thaw depth ranges from 39 cm to 72 cm (*79*).

The model domain represented a two-dimensional groundwater flow transect (Y–Y′ in Fig. 2B) that spanned the hillslope-to-riparian transition zone within the Imnavait Creek watershed. The 109-m long transect (centered at ~ 68°36'39" N, 149°18'56" W) was instrumented with the goal of model-data comparisons. The transect followed a topographic depression identified which forms a water track when overland flow is present, with elevations decreasing 3 m over 109 m from upslope to the Imnavait Creek. Along the transect, 18 piezometers (IMP-01 to IMP-19, Fig. 2B), screened over the bottom 20 cm, were installed. IMP-01 is located at the hillslope break, while IMP-19 is located at the downslope extreme adjacent to Imnavait Creek. Land surface elevations at piezometer locations was surveyed using a robotic laser theodolite (total station) and real-time kinematic (RTK) GPS surveys, yielding elevation measurements with ±1 cm vertical direction accuracy, which were used to define the model's top boundary mesh geometry. Groundwater levels within the aquifer were recorded every 15 minutes using pressure transducers deployed in each piezometer. Thaw or ice table depths were measured manually using a 1.2 m graduated metal probe inserted to refusal, with triplicate measurements taken every ~0.5 m along the transect and averaged to estimate thaw spatial variability. The aquifer's temperature was measured using thermistor arrays (T.rod.X® by Alpha Mach Inc.) of length 50 cm inserted into the ground at four locations, IMP-05, 09, 14 and 18 (Fig. S3). The T.rod.X® combines a data logging controller with six temperature sensors set at different depths within a waterproof PVC housing.

**Input Climate Forcings**

The available long-term hydrometeorological datasets from the Arctic, such as the WERC (Water and Environment Research Center) data (*55*) and Daymet (*58*) have notable limitations. WERC data for the Imnavait Creek location are derived from weather stations and snow measurement platforms requiring high bias corrections and containing discontinuous data. Daymet (*58*) provides meteorological data at 1 km spatial resolution from interpolated observations, but the North slope (~250,000 km$^2$) has very few meteorological stations. Snowfall is estimated using empirical temperature thresholds and precipitation partitioning rules that do not account for processes such as blowing snow, rime deposition, snowpack metamorphism or terrain-informed physics. As a result, both WERC and Daymet often misrepresent snow precipitation and rain-snow partitioning, especially during transitional seasons, and thus they lack spatial correctness.

To address limitations in snow and precipitation records, we used the NASA DAAC ABoVE SnowModel dataset (*19*), which provides a higher resolution, bias-corrected continuous meteorological reconstruction for the Arctic-Boreal region from 1981 to 2020. This dataset integrates multiple observational and reanalysis sources with physically-based snow and energy-balance modeling to account for undercatch, to partition rainfall and snowfall, and to correct biases in snow density and accumulation. Compared to other available products, it offers a spatially and temporally consistent representation of winter precipitation and snowpack evolution, making it well suited for analyzing long-term climate change trends in Arctic Alaska. We further manually calibrated model physics using these forcings to ensure co-temporal alignment with site observations (Fig. S6). Consequent co-temporal and co-spatial calibration was not possible as the forcing data was not available after 2020, and our field observations were collected in the year 2024. We did a co-temporal manual tuning of domain averaged physics from 2010-2020 (Fig. S6). Model output was subsequently compared with field observations from 2024, demonstrating good agreement across key variables and lending confidence to the model's ability to capture important hydrological and thermal trends (Fig. 3).

To understand long term trends in the modeling results, we computed annual means for summer (June-August), winter (September - May) and yearly air temperatures and annual cumulative snow precipitation. To calculate all monotonic trends in this study we applied the Theil-Sen slope estimator (*80*, *81*), a non-parametric stochastic linear regression method that is less sensitive to outliers than ordinary least squares to understand the long term trends.

**Model Geometry**

We used a 2D hillslope - riparian transect (109 m long, 40 m deep, ~2 % slope; Fig. 2C) with coupled surface and subsurface meshes at 1 m horizontal resolution.

Vertical discretization was finest near the surface (1 cm) to resolve dynamic hydrothermal processes and progressively coarsened with depth where the thermal signal is smoother and high resolution is not needed. The organic zone included a thin acrotelm overlying a catotelm, underlain by mineral soil subdivided into shallow fine-resolution cells and coarser cells at depth. This layered structure balanced the need for fine vertical resolution near the surface where thermal and fluid fluxes are most dynamic with coarser resolution at depth to reduce computational cost. Each of the acrotelm, catotelm, and mineral soil zones was assumed to be internally homogeneous and isotropic in terms of hydraulic and thermal properties.

**Boundary Conditions**

The subsurface domain was assigned no-flux boundary conditions for mass and heat on the upslope and downslope sides. The bottom was assumed to remain at a constant temperature of -10 °C, which had been found to be the long-term temperature measured at boreholes near our site (supplementary to Biskaborn et al. (2019), *82*) at an approximate 40 m depth. On the surface , a seepage face boundary condition was applied at the outlet of the downslope end cell (next to the stream, x=109 m) to allow surface water to flow out of the domain. When the local hydraulic head exceeded a prescribed ponded depth, water is allowed to freely leave the domain through the prescription of that threshold head value (0 m); otherwise, a no-flow condition was applied at the outlet boundary. All other boundaries of the surface domain were assigned no-flow conditions, and no-flux conditions were applied for heat across all surface boundaries.

ATS integrates surface energy fluxes, overland thermal flow, and subsurface thermal flow processes (Fig. 2A). Surface energy fluxes balance the incoming and outgoing radiation, sensible heat, conductive heat, and latent heat between the atmosphere and the land surface through meteorological data, providing mass and energy sources to the overland thermal flow process. The surface and subsurface systems are continuous in both fluxes and primary variables (i.e., temperature and pressure). The meteorological data were obtained from the ABoVE SnowModel dataset (*19*), with a daily temporal resolution (1981 to 2020) and a spatial resolution of 3 km × 3 km.

**Model Initialization and Spin-up**

Model initialization and spin-up was conducted with a three-step procedure following previously established routines for permafrost-thermal hydrology modeling (*46*, *50*, *83*). As a first step in model initialization, we performed a one-dimensional (1D) column "freeze-up" simulation to establish a physically consistent initial frozen state for the subsurface domain.

Next, the 1D frozen column was subjected to averaged and smoothed meteorological inputs and run with a complete coupled surface-subsurface flow and energy flux balance to "spin-up" the model. The objective of this step was to bring the thermal and hydraulic condition of the column model into an annual steady state. The annual steady state was achieved by repeating the smoothed and averaged forcing dataset for 100 annual cycles, corresponding to 100 years of simulation, after which the interannual temperature and moisture differences throughout the column are negligible.

In the third step, we performed a transient spin-up of the 1D column model with this spin-up state as initial condition and historical meteorological forcing from 1981 to 2020. We used this additional calibration step to align the model outputs with site-specific observations. Given the low slope of the transect (~ 2 %) , the 1D column model approximately captures some hydrological and thermal dynamics with substantially lower computational time as compared to a full 2D hillslope simulation. We used this efficiency as a first step to calibrate key model parameters, discussed below, against observed data prior to 2D implementation.

Finally, the calibrated pressure and temperature profiles from the 1D column were uniformly mapped to all 109 vertical columns placed horizontally across the 2D transect domain to serve as initial conditions. The full 2D model was then run with real transient meteorological forcing from 1981 to 2020, allowing the system to evolve under historical climate variability and yielding the final model outputs for further analysis.

**Model-Data Comparison and Validation**

We manually calibrated the model using a combination of snow, temperature, and hydrological observations collected along and near the Imnavait Creek transect. As an initial step, the 1D column model, driven by observed meteorological forcing, was evaluated against multi-year observations from the Imnavait Creek SNOTEL station (Site ID: 968, located at 68º37′ N, 149º18′ W; *57*). This site provided records of surface temperature, snow depth, and snow density spanning the period from 2011 to 2020, overlapping with 10 years (25%) of the modeled output.

Initial simulations using default model parameters underestimated wintertime ground insulation, resulting in unrealistically low subsurface temperatures and vertical temperature gradients during the cold season. Resolving this bias is critical, as it could lead to inaccurate initialization of hydro-thermal conditions at the onset of thaw, ultimately affecting seasonal thaw depth and water flux dynamics. Two key factors were identified as potential contributors to this bias: (1) underestimated snow accumulation due to known precipitation undercatch in some Arctic meteorological datasets (*55*, *58*) and (2) the parameterization of snowpack thermal transport. The exponential term in the snow compaction equation was empirically derived based on site-specific conditions. As noted by Martinec (*84*), this exponent could vary with location and prevailing meteorological conditions. Because snow thermal conductivity increased with snow density, higher compaction rates accelerated conductive heat loss from the ground, which resulted in a colder soil temperature.

For calibration, we evaluated multiple combinations of winter snow precipitation bias factors and snow compaction exponents. To adjust for unknown snow undercatch in precipitation measurements, we applied multiplicative scaling factors to winter snow precipitation inputs. Simultaneously, we varied the exponent in the snow densification formulation, which governed the rate of snowpack compaction and thus its thermal conductivity (Supplementary Eq. 31, 32). These parameters were jointly adjusted to optimize agreement between modeled and observed snow depth, snow density, and surface temperature dynamics (Fig. S6) from 2011 to 2020. After iterative calibration, we found that a snow compaction exponent of 0.26 (Eq. 1) instead of the default value 0.3 (Supplementary Eq. 32, *51*), without adjusting the snow precipitation, best fit the model outputs to observations of snow density, snow depth and soil surface temperatures (Fig. S6).

$$\rho_{snow,settled} = \rho_{snow,fresh}(age_{snow})^{0.26} \quad\quad\quad\quad [\text{Eq 1}]$$

Next, we validated the transient two-dimensional (2D) model run using observations of water table depth, thaw depth, snow depth, and soil surface temperature. Modeled outputs from 2011 to 2020 were evaluated against domain averaged field measurements collected along the experimental transect (IMP-01 to IMP-19 in Fig. 2B) during multiple times in the 2024 summer season. The domain-averaged model outputs for water table depth and thaw depth were compared to field observations, with the modeled 10-year mean with a spread of one standard deviation consistently overlapping the observed values across the transect and throughout the thawing season (Fig. 3). Similarly, domain-averaged snowpack evolution and surface temperature from the 2D model were evaluated against daily records from the nearby SNOTEL station (Site ID: 968) from 2011 to 2020. Modeled domain averaged snow depth and soil surface temperature trajectories closely followed the observed seasonal patterns, with mean values and variability well aligned with the SNOTEL sensor data (Fig. 3).

Direct field observations along the 2D transect, including measurements of water table depth, ice table position, and temperature profiles from instrumented groundwater wells at multiple time points during the summer of 2024, showed good agreement with the modeled soil moisture and temperature maps (Fig. 3). The simulated spatial and temporal patterns of the water table, ice table, and temperature agreed with the respective observations (average ± standard deviation) over 10 years and across the transect. The simulated patterns agreed with observations, indicating that the model captures key system dynamics.

**Acknowledgments**

The authors would like to thank Michael O'Connor, Yue Wu, Ke Wang, Jason Dobkowski, Randy Fulweber and the Toolik Field Station for previous data repository and for facilitating our efforts in the field. The Imnavait Creek watershed is part of the hunting grounds and routes of the Nunamiut, Gwich'in, Koyukuk, and Inupiaq people that continue to inhabit and serve as stewards of the land.

**Funding:**

U.S. Department of Energy (DOE), Office of Science, Office of Biological and Environmental Research (BER), Environmental System Science (ESS) Program grant DE-SC0024091 (ETC, PS, BTN, GWK, MBC)

NASA Terrestrial Hydrology Program grant 80NSSC18K0983 (JC, MBC)

Future Investigators in NASA Earth and Space Science and Technology (FINESST) (NM, JC, MBC)

US National Science Foundation grant ARC-1204220 (GWK)
US National Science Foundation grant DEB-1026843 (GWK)
US National Science Foundation grant DEB-1637459 (GWK)
US National Science Foundation grant DEB-0639805 (GWK)
US National Science Foundation grant DEB-2224743 (GWK)
US National Science Foundation grant PLR-1504006 (GWK)
US National Science Foundation grant OPP-1107593 (GWK)
US National Science Foundation grant OPP-1936759 (GWK)

**Author contributions:**
Conceptualization: NM, MBC, ETC, PS, BTN
Methodology: NM, GWK, MBC
Software: NM, BG, ETC, PS
Validation: NM, MBC
Formal Analysis: NM
Investigation: NM, MBC
Resources: NM, JC, MBC
Data Curation: NM
Visualization: NM, MBC
Supervision: BG, ETC, PS, BTN, GWK, JC, MBC
Project administration: ETC, PS, BTN, GWK, JC, MBC
Funding acquisition: ETC, PS, BTN, GWK, JC, MBC
Writing - original draft: NM
Writing - review & editing: BG, ETC, PS, DH, BTN, GWK, JC, MBC

**Competing interests:** Authors declare that they have no competing interests.

**Data and materials availability:**
The source code of the Advanced Terrestrial Simulator (ATS) model are available under the Berkeley Software Distribution (BSD) License at https://github.com/amanzi/ats. All data and model code supporting the findings of this study will be archived in the ESS-DIVE repository and made openly available upon acceptance of the manuscript. Permanent DOIs and access links will be provided in the final published version and in the updated ArXiv.org preprint.

**Figures and Tables**

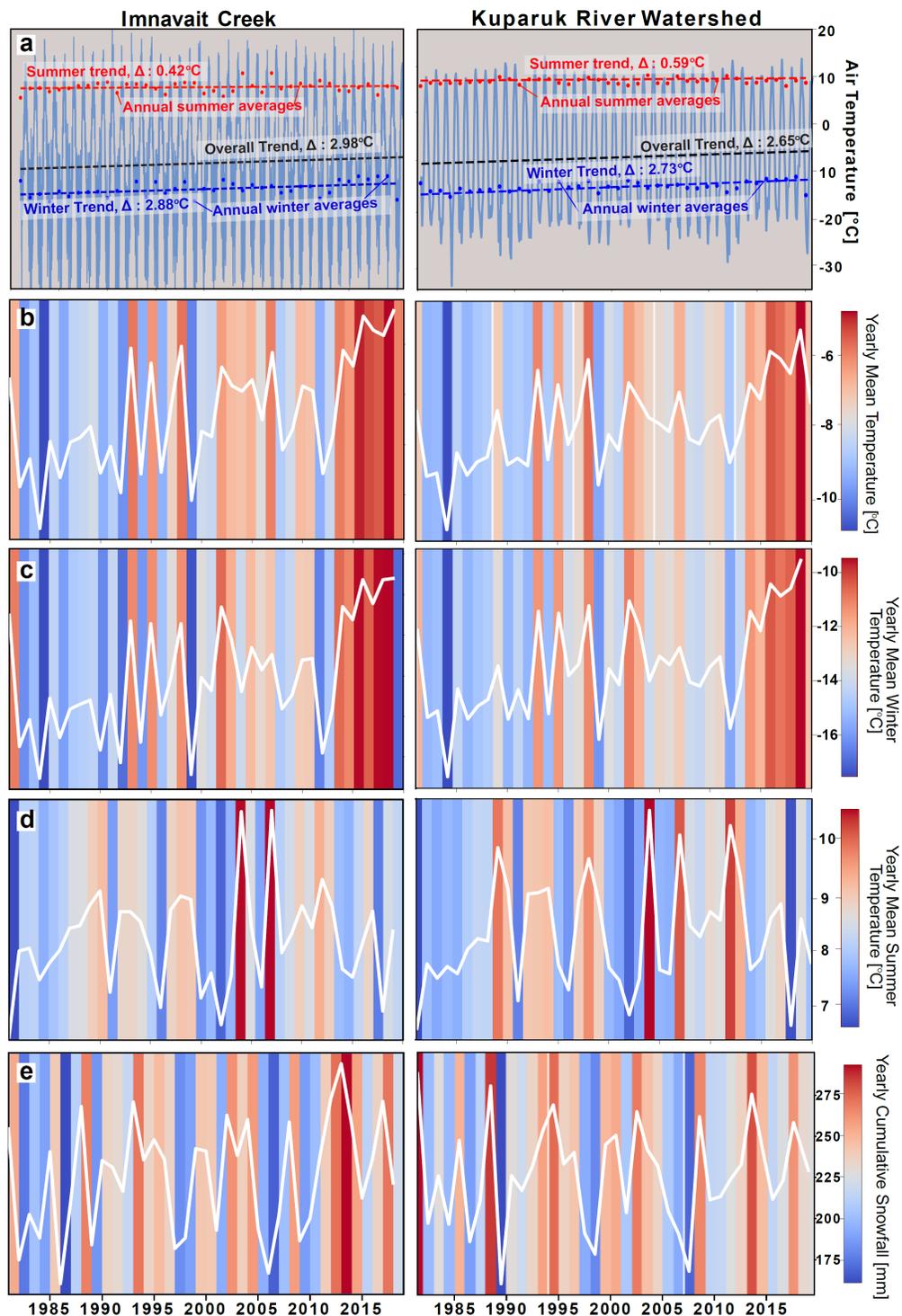

**Fig. 1.** Long-term climate trends at the Imnavait Creek field site and the Kuparuk River watershed, 1981-2020, from the NASA DAAC ABoVE Snowmodel dataset. **(a)** Annual mean, winter (September–May), and summer (June–August) air temperature trends with Theil–Sen trend estimates (Δ), **(b)** Heatmap showing yearly mean temperatures at each site. Each stripe shows average temperature for a single year. Heatmaps showing **(c)** yearly mean winter temperatures for each year, **(d)** yearly mean summer temperatures, and **(e)** yearly cumulative snowfall. All trends are positive over the 40-year period.

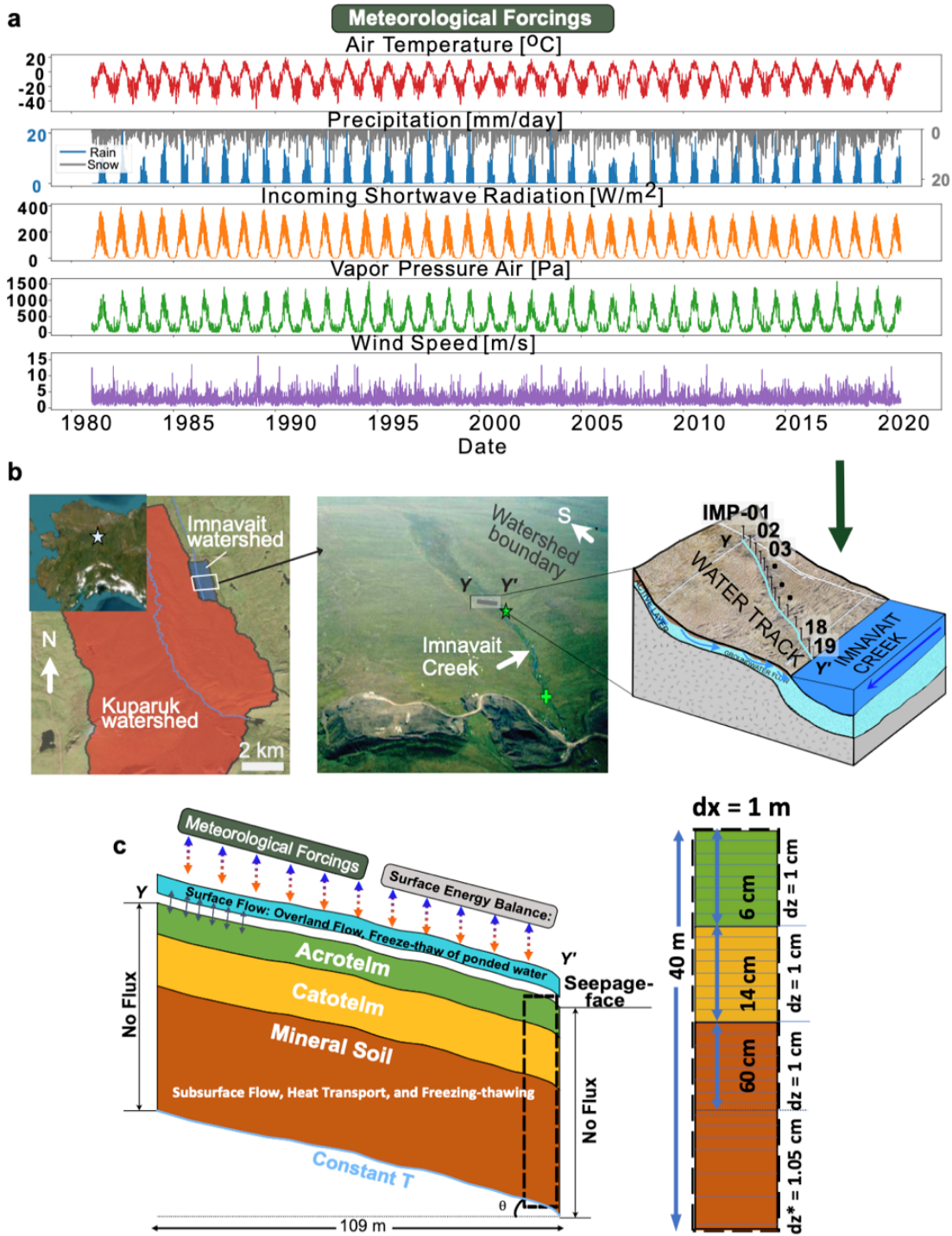

**Fig. 2.** Schematic of Imnavait Creek hillslope-riparian transect and model implementation. **(a)** shows the daily meteorological forcing from NASA DAAC ABoVE Snowmodel Dataset (1981-2020) applied as top boundary condition to the 2D model. **(b)** shows the location of Imnavait Creek watershed and relative location of installed transect of piezometers (IMP-01 to 19), along the 109 m water track (Y-Y') to the Imnavait Creek. **(c)** The left panel shows the coupled processes that are solved in the 2D transect and boundary conditions for the model. The right panel shows the subsurface mesh design showing acrotelm, catotelm and mineral soil layers with vertical resolution refinement near the surface where the supra-permafrost aquifer (active layer) is located.

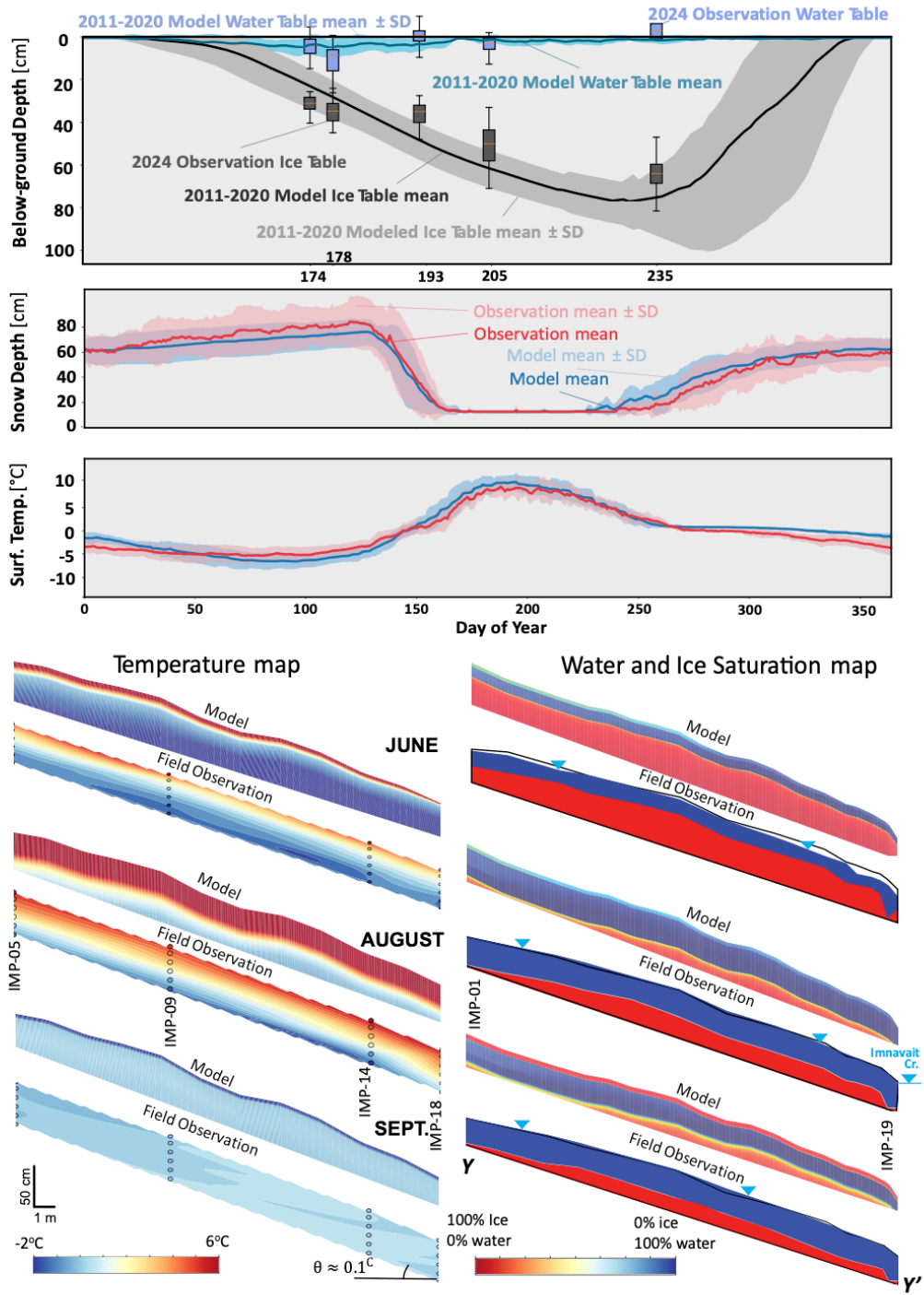

**Fig. 3.** Model-data comparison of snow depth, surface temperature, thaw depth, and water table depth along the Imnavait Creek transect. Top panel: Modeled mean ice table depth and water table depth (2011–2020) compared with 2024 field observations showing good agreement. Domain-averaged snow depth and surface temperature for 2011–2020, comparing model mean ± standard deviation (SD) and observation mean ± SD from the nearby SNOTEL station, which show good agreement. Bottom panel: Modeled temperature profiles and corresponding water and ice saturation maps compared to direct field observations in 2024 for early June, August (peak thaw), and September (freeze-up). The profiles illustrate the model validation with good agreement between model results and field observations.

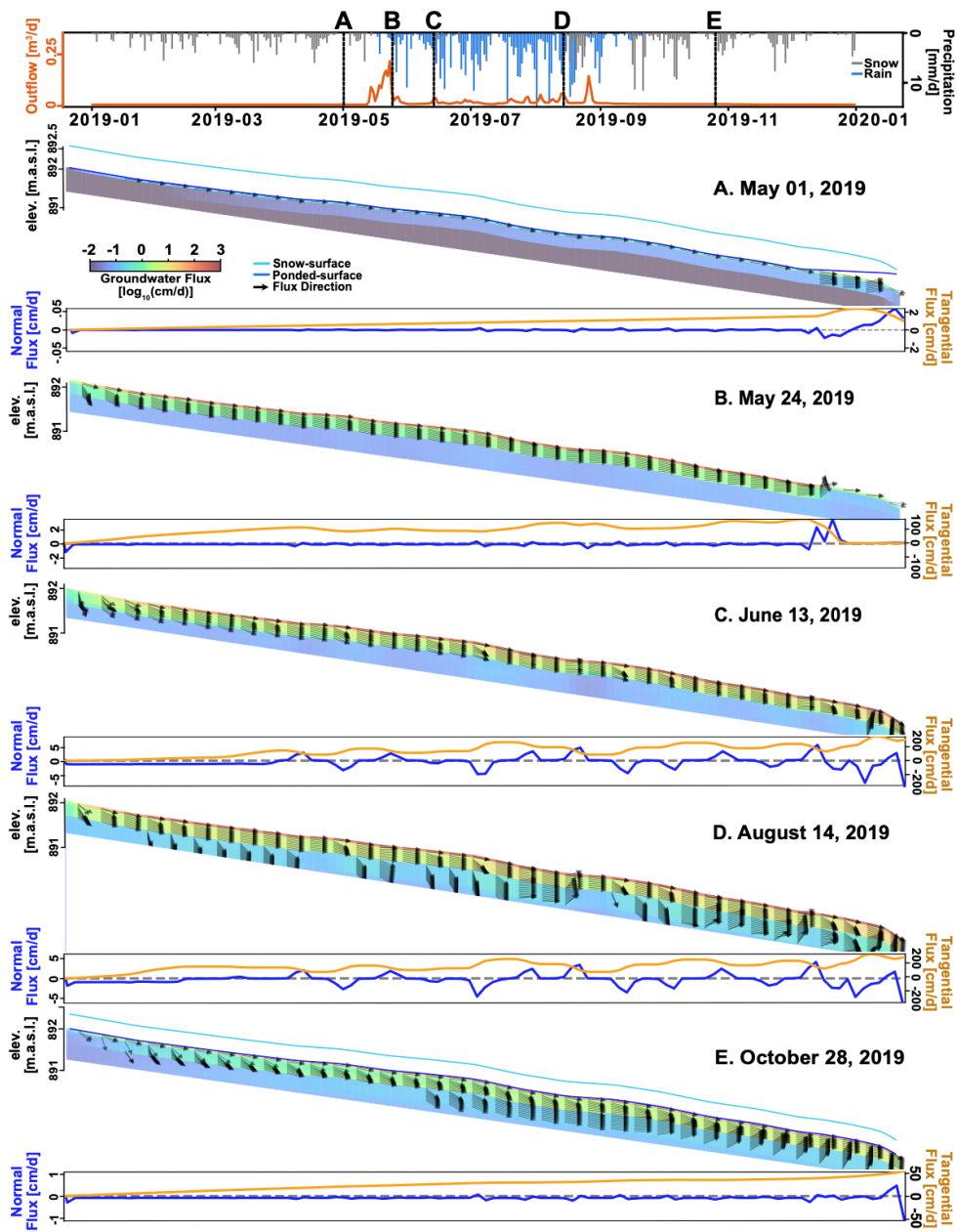

**Fig. 4.** Modeled groundwater flux magnitude and direction along the Imnavait Creek hillslope–riparian transect during representative periods of the 2019–2020 water year: **(a)** May 1 (late winter), **(b)** May 24 (snowmelt/freshet), **(c)** June 13 (early summer thaw), **(d)** August 14 (peak summer), and **(e)** October 28 (freeze-up). Panels show groundwater flux magnitude (in log scale), snow or ponded surface extent (white or blue overlay), groundwater flux directions and normal and tangential fluxes per unit area across the subsurface-surface interface showing flow exchange.

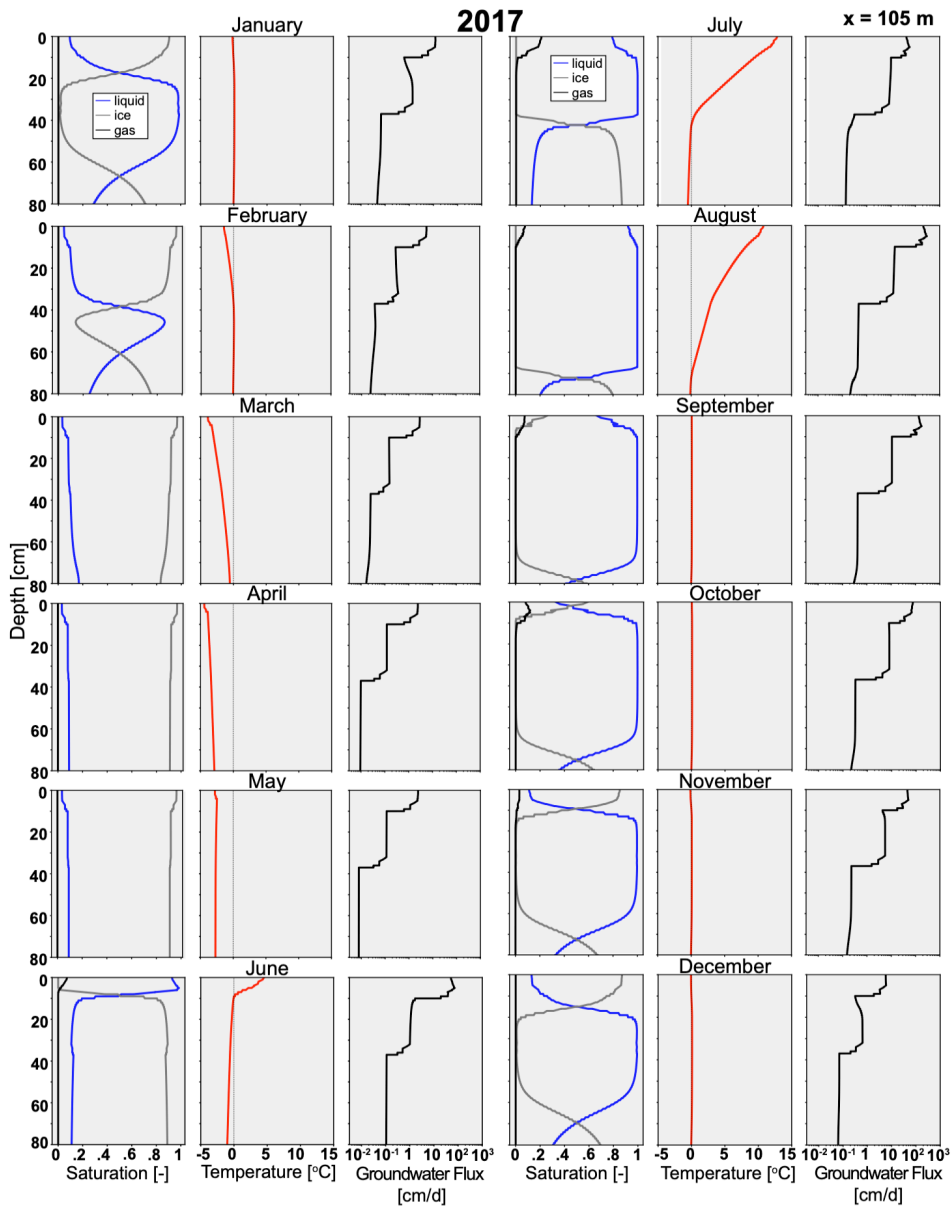

**Fig. 5.** Monthly vertical profiles of liquid water/ice/gas saturation, temperature, and groundwater flux magnitudes at x = 105 m (riparian zone) for 2017. The profiles illustrate seasonal transitions from fully frozen winter (January-March) to partial thaw and snowmelt infiltration (May-June), peak summer saturation and high subsurface flow (July-August), and bidirectional freeze-up with declining flux magnitudes (September-December).

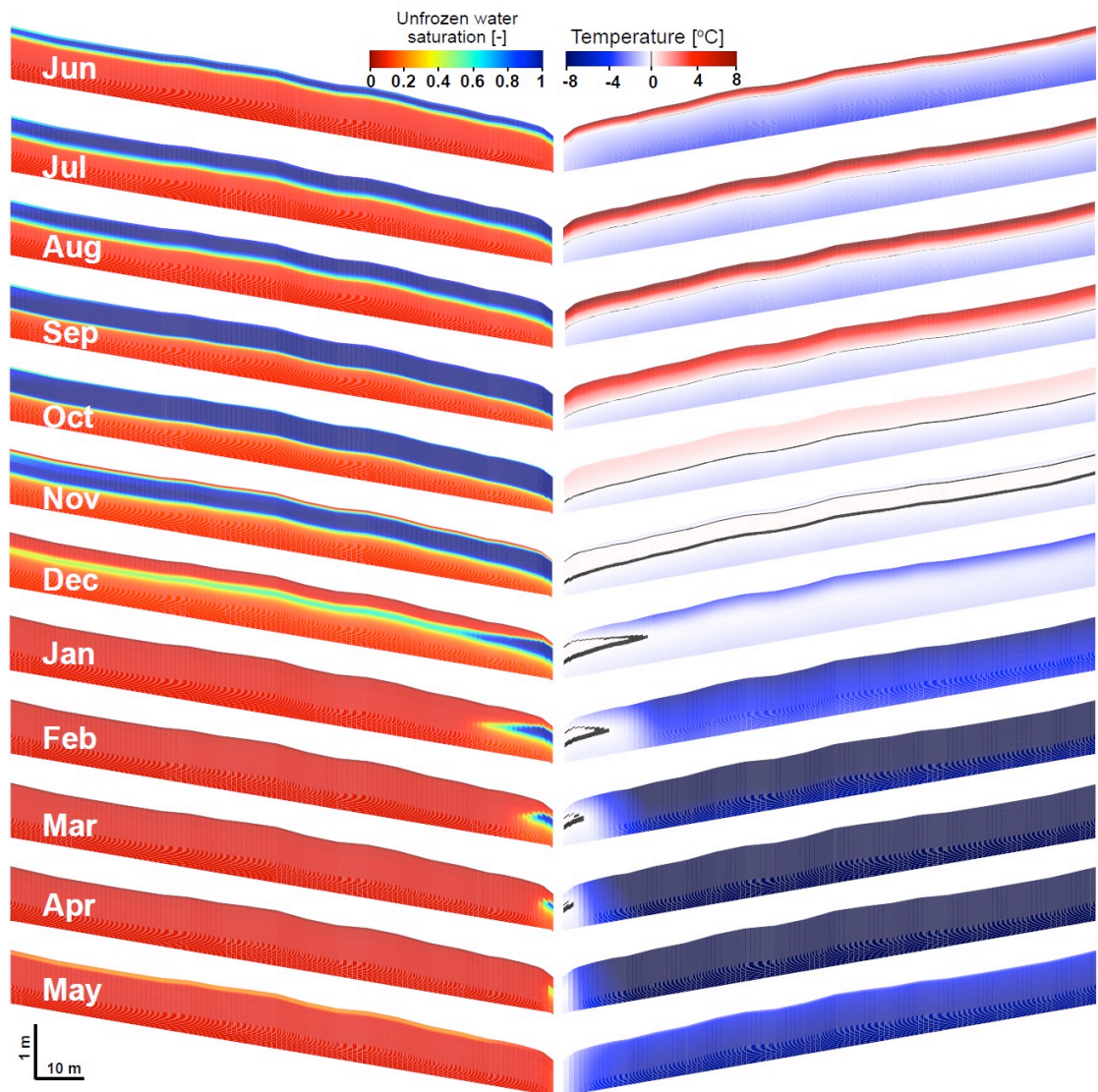

**Fig. 6.** Modeled monthly mean profiles of temperature (left) and unfrozen water saturation (right), averaged across the Imnavait Creek transect for the year 2011. Winter months show near-surface freezing but sustained unfrozen water at depth, especially in the riparian zone. Warmer months (June-August) exhibit full thaw through the active layer, while early winter (September-October) shows bidirectional freezing and the development of zero-curtain conditions.

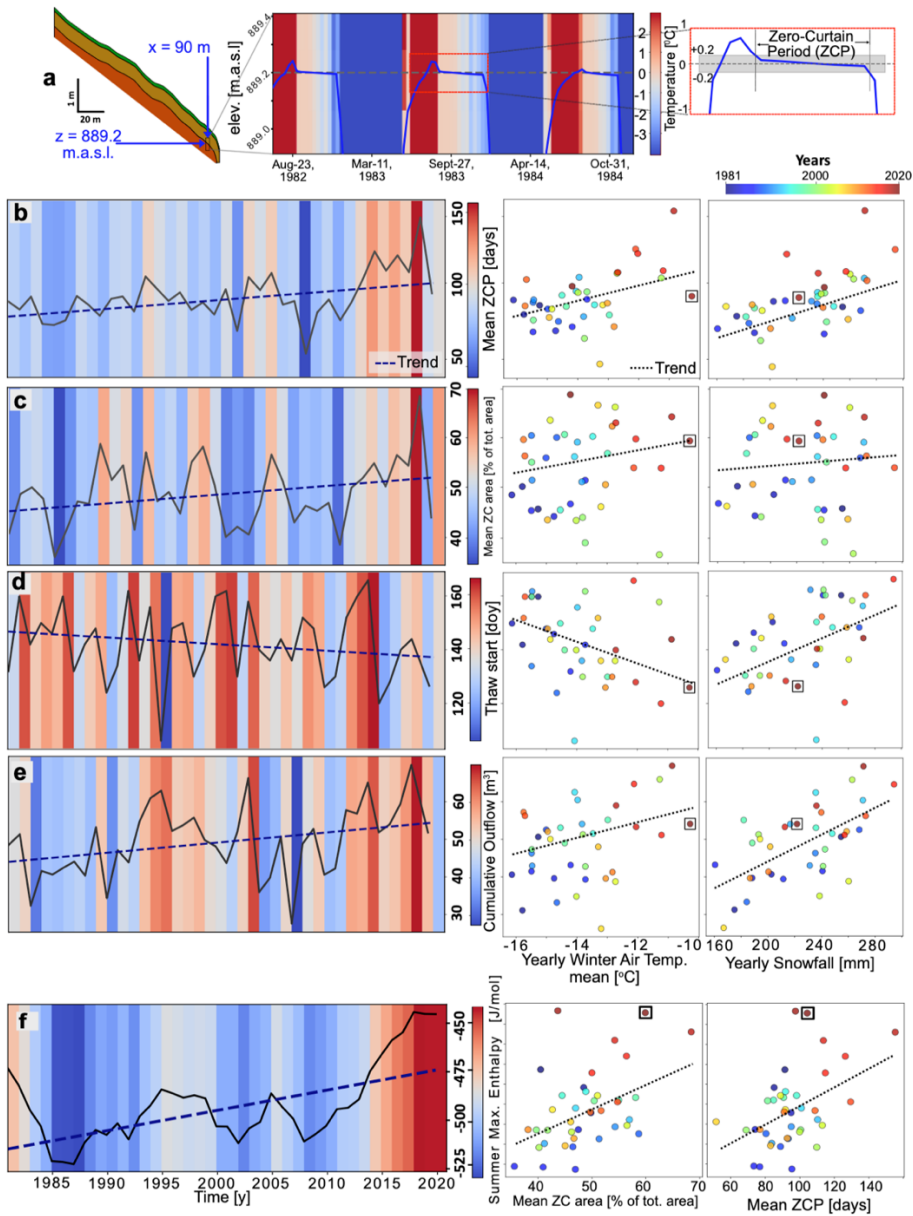

**Fig. 7.** Long-term zero-curtain period (ZCP) dynamics and their relationship with climate drivers along the Imnavait Creek transect. **(a)** Example temperature time series at x = 90 m, z = 889.2 m.a.s.l., showing persistent near 0 °C conditions between thaw onset and complete freeze-up explaining the ZCP. **(b–e)** Left panel shows temporal changes in ZCP duration, mean ZC area as percentage of total unfrozen area, day of year (DOY) when thawing starts, and cumulative yearly outflow, all showing positive trends from 1981 to 2020. Right scatter plots show the Thiel-Sen fitted trends of these yearly metrics with mean winter air temperature and mean annual snowfall from 1981 to 2020. Positive trends indicate lengthening ZCPs, earlier thaw onset, and increasing winter water retention under warmer, snowier conditions. The point corresponding to the warmest winter year (2017 to 2018) is shown with a black square box enclosure. **(f)** Left panel shows the maximum enthalpy at the end of summer per year showing a positive trend from 1981-2020. Right scatter plots show the Thiel-Sen fitted trends of summer maximum enthalpy with mean ZC area and ZCP duration. Both the spatial and temporal winter ZC trends shows a positive relationship with summer maximum enthalpy

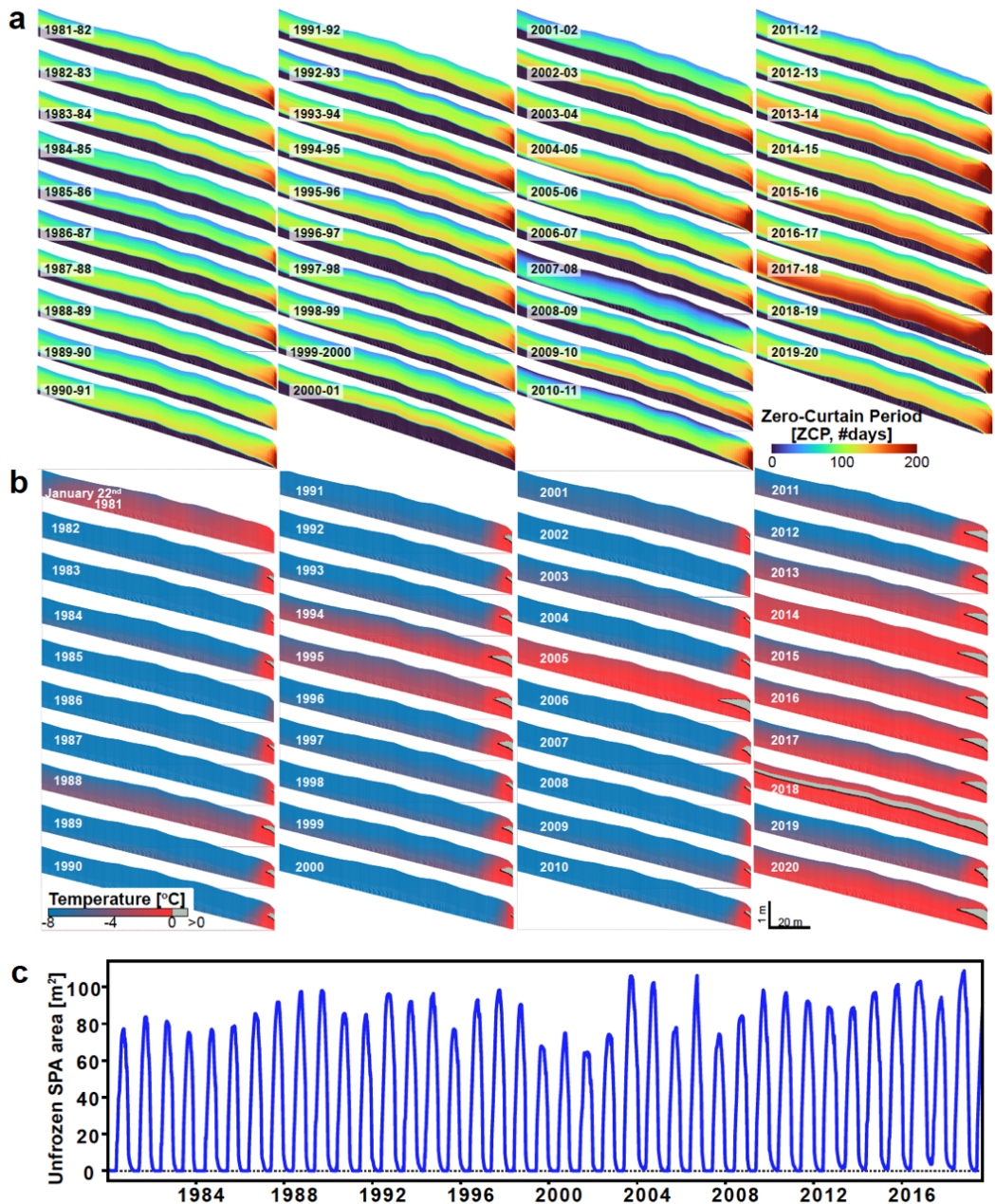

**Fig. 8. (a)** Zero-curtain period for each cell in the domain. The profiles show a consistent increase in ZCP with every year. **(b)** Modeled soil temperature profiles along the Imnavait Creek transect on January 22, the climatological coldest day of the year, for 1981–2020. Profiles highlight the progressive warming of winter soils over four decades, with more recent years exhibiting deeper zones near 0 °C (zero-curtain conditions in gray) and reduced fully-frozen depth, particularly in riparian areas. **(c)** Time series showing total unfrozen area within the supra-permafrost aquifer (SPA). The plot shows an increase in the maximum unfrozen area with time as well as a non-zero unfrozen area in the 2017-2018 winter season when the aquifer did not freeze completely.

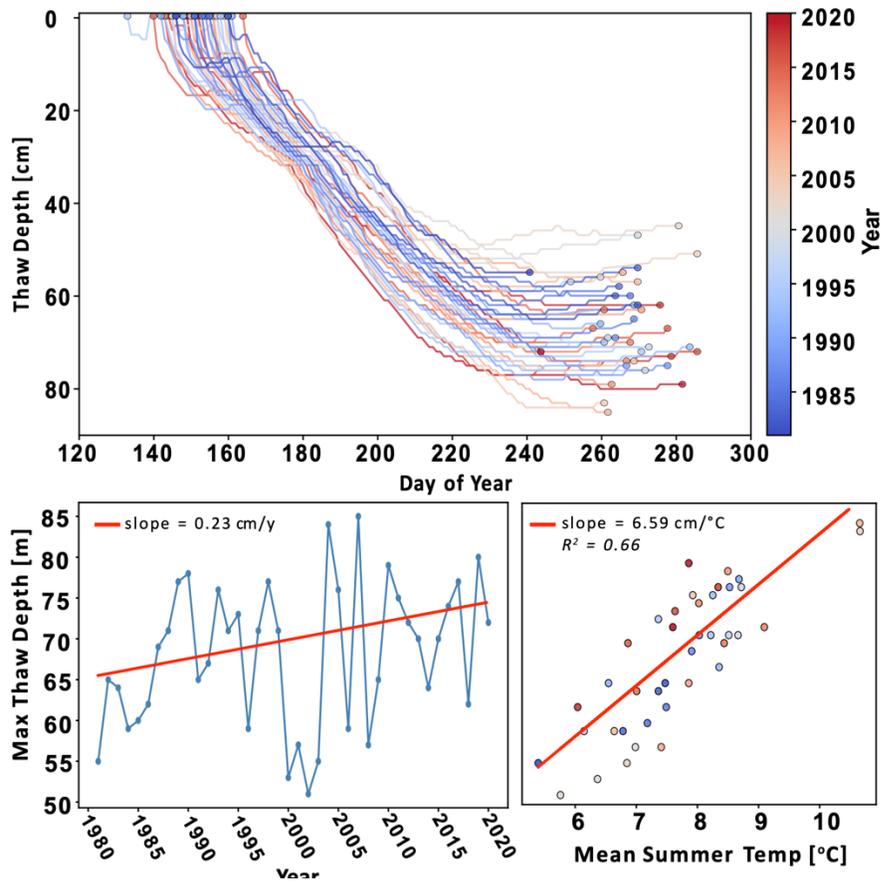

**Fig. 9.** Model results showing thaw depths (ice table) for all the years 1981-2020. Top panel shows the thaw depth time series from day of year 120 to 300. Lower panel shows the maximum thaw depth each year fit with a Thiel-Sen trend on the left, and scatter plot on the right shows the correlation between maximum thaw depth and mean summer (June, July, August) temperature with a Thiel-Sen fit.

Supplementary Materials for

# Climate change impacts on supra-permafrost soil and aquifer hydrology: broader, deeper, and longer activity

Neelarun Mukherjee *et al.*

*Corresponding author. Email: neelarun@utexas.edu

**This PDF file includes:**
    Supplementary Text and References
    Table S1
    Figures S1 to S6
    Caption for Movie S1 and S2

**Other Supplementary Materials for this manuscript include the following:**
    Movie S1
    Movie S2

# Supplementary Text
## Governing Equations for the ATS Model

The physics-based numerical simulations in this study were performed using the Advanced Terrestrial Simulator (ATS, *42*). ATS is a particular construction of the multiphysics software Amanzi (*85*). It is a model for the thermal hydrology of variably-saturated arctic soils and aquifers (*43, 44, 86, 87*). The model structure we used coupled surface and subsurface thermal hydrology with surface energy balance that considers a single-layer model for snow thermal processes. The model includes five "process kernels" (PKs) that each represent a physical process relevant for heat and water flow through the surface above the supra-permafrost aquifer (SPA) and the SPA itself.

A description of the governing equations used in the model is below. All the variable symbols and prescribed input values for parameters are provided in Table S1. The descriptions of surface and subsurface thermal hydrology provided below are also presented in greater detail in Painter (*86*), Painter et al. (*87*), and Karra et al. (*88*), and the descriptions of the surface energy balance are provided in greater detail in Atchley et al. (*51*).

## Subsurface Thermal Hydrology

In order to fulfill conservation of mass and heat in a variably-saturated soil, all phases present in a variably-saturated medium (ice, liquid, and air) must be represented. Painter (*86*), presented a two component (air, water), three-phase (gas, liquid, ice) model with one moving phase (water) in a freezing, variably saturated porous media, and showed that the formulation can reproduce laboratory experiments on freezing soils. Karra et al. (*88*) simplified this by neglecting explicit air conservation and adopted a new phase-partitioning scheme, which is now implemented in ATS. Following this, the conservation of water mass in a transient, variably-saturated soil is governed by the Richards equation, which is expanded here to account for all these phases and expressed in molar form:

$$\frac{\partial}{\partial t}\left[\varphi\left(\sum_{P=l,i,g} \omega_P \eta_P s_P\right)\right] = -\nabla \cdot \left[\eta_l \vec{V_l}\right] + q_W, \quad \text{[Eq. 1]}$$

where $\varphi$ is porosity [-], $P$ represents substance phase (with $l, g$ and $i$ represent liquid, gas and ice phases respectively), $\omega$ is mole fraction [-], $\eta$ is molar density [mol m$^{-3}$], $q_w$ is a fluid source [mol m$^{-3}$ s$^{-1}$], and $s$ is the phase saturation ($s_g + s_l + s_i = 1$). As both ice and liquid water are assumed to be pure-component phases (i.e., both ice-phase air and dissolved air are ignored), $\omega_l = 1$ and $\omega_i = 1$. $\omega_g$ is defined as:

$$\omega_g = \frac{e_{sat}\{T\}}{p_g}, \quad \text{[Eq. 2]}$$

where $e_{sat}\{T\}$ is the saturation vapor pressure at a given temperature $T$ [Pa], calculated from the Clausius-Clapeyron Equation, and $p_g$ is the gas partial pressure, assumed to be 101325 Pa.

ATS assumes that the change of mass storage is governed only by the movement of liquid phase water (*83*). The groundwater flux (also referred to as the Darcy velocity) $\vec{V_l}$ [m s$^{-1}$] is therefore defined through Darcy's Law:

$$\vec{V_l} = -\frac{k_{r,l}k}{\mu_l\{T\}}(\nabla p_l + \rho_l\{T\}g\hat{z}), \quad \text{[Eq. 3]}$$

where $k_{r,l}$ is the relative permeability [-], $k$ is intrinsic permeability [m²], $\mu_l\{T\}$ is liquid water dynamic viscosity [Pa s], $p_l$ is pore pressure [Pa], $\rho_l\{T\}$ is liquid water mass density [kg m⁻³], g is acceleration due to gravity [m s⁻²], and $\hat{z}$ is the z-dimension unit vector [m].

The water mass balance equation is tightly coupled to the energy balance equation for unsaturated soil, which is constructed assuming local thermal equilibrium between all phases:

$$\frac{\partial}{\partial t}\left[\varphi \sum_{p=l,g,i}(\eta_p s_p u_p) + (1-\varphi)C_{v,soil}T\right] = -\nabla \cdot (\eta_l h_l \vec{V_l}) + \nabla \cdot (\kappa_{eff}\nabla T) + Q_{g-gw} + Q_{Ess}$$
[Eq. 4]

where $u_l$ is the specific internal energy of liquid water [J mol⁻¹], $u_i$ is the specific internal energy of ice [J m⁻³], $u_g$ is the specific internal energy of gas [J mol⁻¹], $C_{v,soil}$ is the volumetric heat capacity of the soil matrix [J m⁻³ K⁻¹], $h_l$ is the specific enthalpy of liquid water [J mol⁻¹], $\kappa_{eff}$ is effective thermal conductivity [W m⁻¹ K⁻¹], $Q_{g-gw}$ is the convected heat source from the ground [W m⁻³], and $Q_{Ess}$ is the conducted heat source from the ground [W m⁻³].

We represent the specific internal energy of liquid water as a linear function of temperature about the melting point. The internal energy of liquid water is given as

$$u_l(T) = C_l(T - T_{ref}),$$
[Eq. 5]

where $C_l$ is the constant molar heat capacity of liquid water ($C_l = 76$ J mol⁻¹ K⁻¹), and $T_{ref} = 273.15\ K$ is the reference temperature corresponding to the phase transition between liquid water and ice. We calculated the specific enthalpy of the domain, $h_l$ to track heat stored in the domain, defined as

$$h_l = u_l + \frac{p}{\rho_l},$$
[Eq. 6]

$p$ is pressure and $\rho_l$ is mass density of liquid water. $\frac{p}{\rho_l}$ ratio is generally very small compared to internal energy associated with phase change, heating/cooling in saturated soils, we neglect the pressure-volume work term and approximate $h_l = u_l$. The enthalpy is negative when $T < T_{ref}$, because the specific internal energy, $u_l$, is negative below $T_{ref} = 273.15\ K$. Along the model-transect, subsurface temperatures remain below freezing from roughly 1 m to 40 m depth (below maximum active layer depth). Thus, negative enthalpy values simply indicate that the subsurface has less internal energy than the liquid-water reference state.

To calculate the domain-averaged enthalpy, we compute a volume averaging over the subsurface domain. Let $h_k$ denote the specific enthalpy in cell $k$ and $V_k$ is the volume of the cell. The domain averaged enthalpy (Fig. S5) is then defined as

$$\bar{h} = \frac{\sum_k h_k V_k}{\sum_k V_k}$$
[Eq. 7]

Unlike in the water mass balance, conduction and convection of heat from multiple phases are considered in the total energy balance. The dependent variables in the system of equations created by Eq. 1, 3, and 4 are $p$ and $T$. Tight couplings exist between the thermal and water mass balances due to changing saturation indices s and fluid advection $V_l$.

The inclusion of phase change into thermal hydrology models introduces additional tight couplings between the water and energy balances. Freeze/thaw in porous media is governed by the capillary forces between the phases present: ice, water, and air (*87*). Tensile capillary forces between air, ice, and water can depress the freezing point temperature ($T_f$) in porous media, because water molecules that are more tightly-adhered to the soil matrix due to capillary forces will require more energy to change phase (*48*). Thus, the partitioning between $s_l$, $s_i$, and $s_g$ is a function of both temperature and pressure. To account for this additional coupling, ATS employs an explicit formulation of the Clapeyron equation, which describes the relationship between temperature, capillary pressure, and freezing point depression in variably-saturated soils except for tight clays (*89*):

$$s_l = \begin{cases} s_*\{-\beta \rho_l L_f \vartheta\}, & \vartheta < \vartheta_f \\ s_*\{p_g - p_l\}, & \vartheta \geq \vartheta_f \end{cases}, \quad \text{[Eq. 8]}$$

$$\vartheta_f = -\frac{1}{\beta L_f \rho_l} \psi_*\{1 - s_g\}, \quad \text{[Eq. 9]}$$

$$s_i = 1 - \frac{s_l}{s_*\{p_g - p_l\}}, \quad \text{[Eq. 10]}$$

where $L_f$ is the latent heat of fusion of water [J kg$^{-1}$], $\vartheta$ [-] is a dimensionless temperature $\left(\frac{T-T_0}{T_0}\right)$, $p_g$ is the capillary pressure of the gas phase [Pa], and $p_l$ is the capillary pressure of the liquid phase [Pa]. $\beta$ [-] is a coefficient related to soil quality: $\beta = 1$ for colloidal soils (i.e., granular soils with substantial grain-to-grain contact), and $\beta = \left(\frac{\gamma_{i-l}}{\gamma_{l-g}}\right)$ for noncolloidal soils (i.e., soils with substantial grain-water-grain contacts), where $\gamma_{i-l}$ is the surface tension between ice and liquid phases, and $\gamma_{l-g}$ is the surface tension between liquid and gas phases (*90*).

Any coupling between soil capillary pressures and saturation states requires a constitutive relationship between pressure and saturation; in hydrology, these relationships are most commonly reported as soil moisture retention curves. Soil moisture retention curves are unique for individual soils, and many models have been developed to empirically describe the soil moisture retention curve as a function of soil properties. ATS employs the van Genuchten soil moisture retention model (*91*) both because it is widely-accepted in the literature and produces a continuous, differentiable soil moisture retention curve. It is presented here in both its forms (saturation as a function of pressure and pressure as a function of saturation):

$$s_*\{p\} = s_r + (1 - s_r)[1 + (\alpha p)^n]^m, \quad \text{[Eq. 11]}$$

$$\psi_*\{s\} = \frac{1}{\alpha}\left[\left(\frac{s-s_r}{1-s_r}\right)^{-1/m} - 1\right]^{1/n}, \quad \text{[Eq. 12]}$$

where $m = 1 - 1/n$, and $\alpha$ [m$^{-1}$], n [-], and $s_r$ [-] are all either experimentally derived or assumed parameters based on soil texture. In current thermal hydrology models, it is assumed that the capillary forces that adhere porewater to grains in a drying, unsaturated soil are analogous to capillary forces that adhere liquid porewater to grains as an unsaturated soil freezes (*92*). Thus, the soil moisture retention curve, which describes how capillary forces affect the drying of a soil, is assumed to be analogous to the 'soil freezing curve', which describes how capillary forces affect the freezing of the soil (*92*).

The freezing-as-drying approximation is also employed to determine the relative permeability $k_r$ of a variably-saturated porous medium. The permeability of frozen soil is mainly controlled by air-filled micropores and the ice content in the soil. A constitutive relation that uses the van Genuchten and Mualem models:

$$k_r = (s_l)^{1/2} \left[1 - \left(1 - (s_l)^{\frac{1}{m}}\right)^m\right]^2 \qquad [Eq.13]$$

However, this constitutive relation is not a function of ice saturation in a cell, which may lead to situations where a frozen ground will not allow infiltration of snowmelt or rainfall resulting in a drier subsurface and a higher than observed runoff. To avoid this problem, ATS has the option to use a modified Brooks-Corey water retention model following the formulation implemented by Niu & Yang (*93*). There are other studies (e.g., *94, 95*) that parametrize frozen soil permeability, however, Agnihotri et al. (*96*) showed that the Niu & Yang (*93*) formulation produces more accurate streamflow simulations in frozen ground dominated drainage areas. In this case, the permeability of the soil has been increased using an ice-impedance coefficient, $F_{frz}$, a function of ice saturation.

$$k_r = (1 - F_{frz}) \left(\frac{s_i + s_l - s_r}{1 - s_r}\right)^{\frac{2}{\lambda} + 3} \qquad [Eq.\ 14]$$

$$F_{frz} = e^{-\varsigma \left(1 - \frac{s_i - s_r}{1 - s_r}\right)} - e^{-\varsigma} \qquad [Eq.15]$$

Where $\lambda$[-] is pore size distribution index for a soil-type and is determined experimentally, and $\varsigma$ [-] is a fitting parameter. ATS calculates the effective thermal conductivity of each cell $\kappa_{eff}$ based on a ratio of saturated, dry, frozen, and unfrozen phases:

$$\kappa_{eff} = Ke_f * \kappa_{sat} + Ke_u * \kappa_{sat} + (1 - Ke_f - Ke_u \kappa_{dry}) \qquad [Eq.\ 16]$$

where $\kappa_{sat}$ is the thermal conductivity of the saturated soil [W m$^{-1}$ K$^{-1}$], $\kappa_{dry}$ is the thermal conductivity of the dry soil [W m$^{-1}$ K$^{-1}$], and the Kersten number Ke is a ratio between partially saturated to fully saturated thermal conductivity [-], calculated as:

$$Ke_u = s_l^{\tau_u} \qquad [Eq.\ 17]$$

$$Ke_f = s_i^{\tau_f} \qquad [Eq.\ 18]$$

where $\tau$ is an empirical fitting parameter that varies for the frozen and unfrozen states [-] (*88*).

**Ground surface coupled water and energy balance including phase change**

ATS employs a system of energy balance equations to allow for 'icy overland flow':

$$\frac{\partial}{\partial t}[(\chi \eta_l + (1-\chi)\eta_i)d_w] + \nabla \cdot (\chi \eta_l d_w \vec{U_w}) = q_{rain} + q_{melt} - q_{evap} - q_{g-gw} \qquad [Eq.\ 19]$$

$$\vec{U_w} = -\frac{(\chi d_w)^{\frac{2}{3}}}{N(\|\nabla z_s\| + \delta)^{\frac{1}{2}}} \nabla(z_s + d_w) \qquad [Eq.\ 20]$$

$$\frac{\partial}{\partial t}[(\chi \eta_l u_l + (1-\chi)\eta_i u_i)d_w] + \nabla \cdot (h_l \chi \eta_l d_w \vec{U_w}) - \nabla \cdot [(\chi \kappa_l + (1-\chi)\kappa_i)d_w \nabla T_s] = Q_{s-g} + Q_{rain} + Q_{melt} - Q_{evap} - Q_{g-gw} - Q_{Ess} \qquad [Eq.\ 21]$$

where $h_l$ is the specific enthalpy of liquid water [J mol$^{-1}$], N is the Manning's Roughness coefficient [s m$^{-1/3}$], $\delta$ is a numerical surface velocity fitting parameter [-], $q$ represents mass fluxes of water [mol s$^{-1}$ m$^{-2}$] and $Q$ represents energy fluxes to the surface [W m$^{-2}$]. The energy balance of the surface cell (Eq. 19) depends on both the mass and the energy fluxes of water sources from above and below, because the total heat flux $Q$ is the product of the mass flux in moles, $q$, and the enthalpy of the system $h$, which depends on temperature.

The mass balance component solves for ponded water depth ($d_w$) using a hybrid of the kinematic and diffusion wave approximations (Eq. 19-20). The hybrid form both approximates the diffusion wave equation on sloping ground and provides physically reasonable results on flat ground. The fluid source term that couples surface water flow to groundwater flow, $q_{s-gw}$, is determined by employing the mass-conservative boundary condition strategy described in Kollet & Maxwell (*97*). Surface and subsurface are coupled by the residual source term $q_{g-gw}$ of equation 16 as a Neumann boundary condition of subsurface face and pressure continuity.

The energy balance component (Eq. 19) includes the advection of heat due to lateral surface water flow $\vec{U_w}$. It also includes phase change through the inclusion of a liquid-ice partitioning factor $\chi$. This factor is represented as smoothed step function with a prescribed width $\tau$ [K] that empirically assigns liquid and ice fractions based solely on $T_s$. This approximation is appropriate for shallow water depths (i.e., a well-mixed water column with no substantial vertical stratification in surface water temperature distributions) but fails in deeper water columns as it does not allow for depth-variable freezing. In this system of equations, water mass is represented in molar form, as this allows for the volumetric expansion of water as it freezes from liquid to ice.

**Surface energy balance, including snowmelt**

The surface energy balance employed in ATS, in a general form, is:

$$Q_{s-g} = Q_{sw}^{in} + Q_{lw}^{in} + Q_{lw}^{out}\{T\} + Q_h\{T\} + Q_e\{T\} \qquad \text{[Eq. 22]}$$

$$Q_{s-g} = Q_{Ess} + Q_c \qquad \text{[Eq. 23]}$$

The right-hand side of this energy balance represents incoming shortwave and longwave radiation $Q_{sw}^{in} + Q_{lw}^{in}$ (provided by meteorological forcing) and outgoing longwave radiation $Q_{lw}^{out}\{T\}$, sensible heat flux $Q_h\{T\}$, and latent heat flux $Q_e\{T\}$ (calculated based on temperature, see below). All energy fluxes presented in Equations 19 and 20 are in [W m$^{-2}$].

The left-hand side of this energy balance, $Q_{s-g}$, represents heat transferred from the surface to the ground. $Q_{s-g}$ is the sum of bare-ground conductive heat flux ($Q_{Ess}$) and snowpack conductive heat flux ($Q_c$) (Eq. 23). ATS assumes that if a snowpack is present, it is in equilibrium with all energy fluxes going into and out of the snowpack; therefore, the term representing residual heat conducted from the surface to the ground, $Q_{Ess} = 0$. This does not mean that heat transfer between the surface and the ground cannot occur, but such heat transfer is controlled by the conduction of heat through the snowpack $Q_c$, which ATS directly calculates based on the thickness, density, and temperature of the snowpack (see below). Conversely, if a snowpack is absent, conductive snowpack heat transfer does not occur, so $Q_c = 0$. Without a snowpack, no thermal equilibrium can be assumed, and $Q_{Ess}$ is calculated as the residual of the right-hand-side energy balance terms. Snowpack conduction $Q_c$ is important for the accurate representation of arctic thermal hydrology, as it can be a controlling factor in ground temperature and SPA development (*51, 98*).

The radiative components of the surface energy balance (incoming $Q_{sw}$ and $Q_{lw}$) are supplied as forcing data. The initial $Q_{sw}$ supplied by the user is corrected to account for ground surface albedo $\alpha$ [-]. Given that a changes with surface temperature due to the shifting exposure of snow, ponded water, or bare ground, a is assigned based on the current surface conditions.

Outgoing longwave radiation $Q_{lw}^{out}$ is calculated based on snow or surface temperature $T_s$:

$$Q_{lw}^{out} = -\varepsilon_s \epsilon T_s^4 \qquad \text{[Eq. 24]}$$

where $\varepsilon_s$, surface emissivity [-], varies with ground cover as a does and $\epsilon$ is the Stefan-Boltzman constant [W m$^{-2}$ K$^{-4}$].

The non-radiative components (latent heat $Q_e$, sensible heat $Q_h$, and conduction $Q_c$ or $Q_{gf}$) are dependent on soil or snow temperature $T_s$. $Q_h$ is expressed as:

$$Q_h = \rho_g C_p D_{eh} \xi (T_a - T_s), \qquad \text{[Eq. 25]}$$

where $\rho_g$ is the mass density of gas [kg m$^{-3}$] and $C_p$ is the specific heat of water [J kg$^{-1}$ K$^{-1}$]. $T_a$ is provided from meteorological forcing. $D_{eh}$, the turbulent exchange of latent and sensible heat, is:

$$D_{eh} = \frac{0.1681 U_s}{(\ln(\frac{z_r}{z_0}))^2}, \qquad \text{[Eq. 26]}$$

where wind speed $U_s$ [m s$^{-1}$] is provided as forcing data, $z_r$ is the reference height of the wind speed measurement [m], and $z_0$ is a prescribed roughness length [L]. The stability function $\xi$ from Equation 22 is:

$$\xi = \begin{cases} \frac{1}{1+10 R_i}, & T_s \leq T_a \\ 1 - 10 R_i, & T_s > T_a \end{cases} \qquad \text{[Eq. 27]}$$

which itself depends on the atmospheric stability parameter $R_i$:

$$R_i = \frac{g z_r (T_a - T_s)}{T_a U_s^2} \qquad \text{[Eq. 28]}$$

Latent heat $Q_e$ is calculated slightly differently depending on if there is a snowpack present or absent. The porosity of the top subsurface cell is used when no snowpack exists, whereas with a snowpack, there is no subsurface influence:

$$Q_e = \begin{cases} \rho_g \lambda_s E_r \left(0.622 \frac{e_a - e_s}{p_{atm}}\right), & z_s \geq 2\ cm \\ \varphi_{z=0} \rho_g \lambda_e E_r \left(0.622 \frac{e_a - e_s}{p_{atm}}\right), & z_s < 2\ cm \end{cases}, \qquad \text{[Eq. 29]}$$

where $\lambda_s$ is the latent heat of sublimation for snow [J kg$^{-1}$], $e_a$ and $e_s$ are the vapor pressures of the atmosphere and the snow or soil, respectively [Pa], $p_{atm}$ is atmospheric pressure [Pa], which is provided as forcing data, and $\lambda_e$ is the latent heat of evaporation from the ground surface [J kg$^{-1}$]. Evaporation resistance $E_r$ [m s$^{-1}$] is the inverse of the air and soil resistance sums $R_{air}$ [s m$^{-1}$] and $R_{soil}$ [s m$^{-1}$], defined as:

$$E_r = \frac{1}{R_{air} + R_{soil}}. \qquad \text{[Eq. 30]}$$

Air resistance is the inverse of turbulent exchange (Eq. 26) and atmospheric stability (Eq. 27):

$$R_{air} = \frac{1}{D_{eh} \xi} \qquad \text{[Eq. 31]}$$

Soil resistance $R_{soil}$ is given as (99):

$$R_{soil} = \exp(8.206 - 4.255 s_l) \qquad \text{[Eq. 32]}$$

ATS solves for conductive heat flux through the snowpack when it exists:

$$Q_c = \begin{cases} \frac{\kappa_s(T_s - T_{ground})}{z_s}, & z_s \geq 2\ cm \\ 0, & z_s < 2\ cm \end{cases} \quad [Eq.\ 33]$$

The thermal conductivity of the snowpack $\kappa_s$ is calculated empirically through the function established in Ling & Zhang (2004, *100*):

$$\kappa_s = 2.9 * 10^{-6} \rho_{snow}^2 \quad [Eq.\ 34]$$

and the total mass density of the snowpack $\rho_{snow}$ [kg m$^{-3}$] is the volume-weighted average of fresh snow ($\rho_{snow,fresh}$, assigned at 100 kg m$^{-3}$), ice from condensation ($\rho_{snow,cond.}$, assigned at 200 kg m$^{-3}$), and older snowpack that settles over time ($\rho_{snow,settled}$), which is empirically calculated based on snow age (*84*):

$$\rho_{snow,settled} = \rho_{snow,fresh}(age_{snow})^{0.3} \quad [Eq.\ 35]$$

Martinec (*84*) noted that the snow compaction exponent, 0.3, is conditional to fresh snow conditions, location, instantaneous relative humidity and the time period when the snow was collected. The total snow density is then used with the cell dimensions to calculate the snow height $z_s$. With snow height and density, ATS then calculates snow water equivalent (SWE):

$$SWE = \frac{z_s}{\rho_{snow}} \quad [Eq.\ 36]$$

As both water fluxes and energy fluxes are necessary to solve the ground energy and mass balance, ATS calculates water fluxes from the surface to the subsurface after calculating the energy balance. ATS iterates to solve for the meltwater contribution to the system: if the energy balance calculation determines that snow temperature $T_s$ is above freezing, then ATS sets $T_s$ to freezing (273.15 K) and the surface energy balance is re-calculated. In this case, all the excess energy that would have gone to snow temperature is then redistributed to snowmelt energy $Q_{melt}$. This allows for ATS to calculate the flux of snowmelt into the subsurface:

$$q_{melt} = \frac{Q_{melt}}{\rho_l L_f} \left( \frac{\rho_w GFW_w}{1000} \right) \quad [Eq.\ 37]$$

Water loss due to both evaporation and sublimation from the snowpack is calculated based on the latent heat flux $Q_e$:

$$q_{evap} = \frac{Q_e}{\rho_l \lambda_s} \left( \frac{\rho_w GFW_w}{1000} \right) \quad [Eq.\ 38]$$

In both equations 37 and 38, the conversion factor $\left( \frac{\rho_w GFW_w}{1000} \right)$ is applied so that the mass fluxes of water are expressed in [mol m$^{-2}$ s$^{-1}$].

**Supplementary Tables**

**Table S1. ATS Model Input Parameters**

| Property | Acrotelm | Catotelm | Mineral | Units |
|---|---|---|---|---|
| Base porosity | 0.88 | 0.8 | 0.46 | - |
| Pore compressibility | 1E-07 | 1E-07 | 1E-08 | Pa$^{-1}$ |
| Permeability | 1.29E-10 | 4.72E-12 | 1.82E-13 | m$^2$ |
| Rock density | 1000 | 1500 | 2000 | Kg/m$^3$ |
| Thermal conductivity (sat, unfrozen) | 0.52 | 0.63 | 1.30 | W/(m-K) |
| Thermal conductivity (dry) | 0.07 | 0.09 | 0.76 | W/(m-K) |
| van Genuchten alpha | 0.000793 | 0.000175 | 8.03E-05 | Pa$^{-1}$ |
| van Genuchten n | 1.40 | 1.57 | 1.57 | - |
| Residual saturation | 0.007 | 0.066 | 0.066 | - |
| Brooks-Corey lambda | 0.37 | 0.49 | 0.49 | - |
| Brooks-Corey saturated matric suction | 892 | 3728 | 8105 | Pa |

**Supplementary Figures**

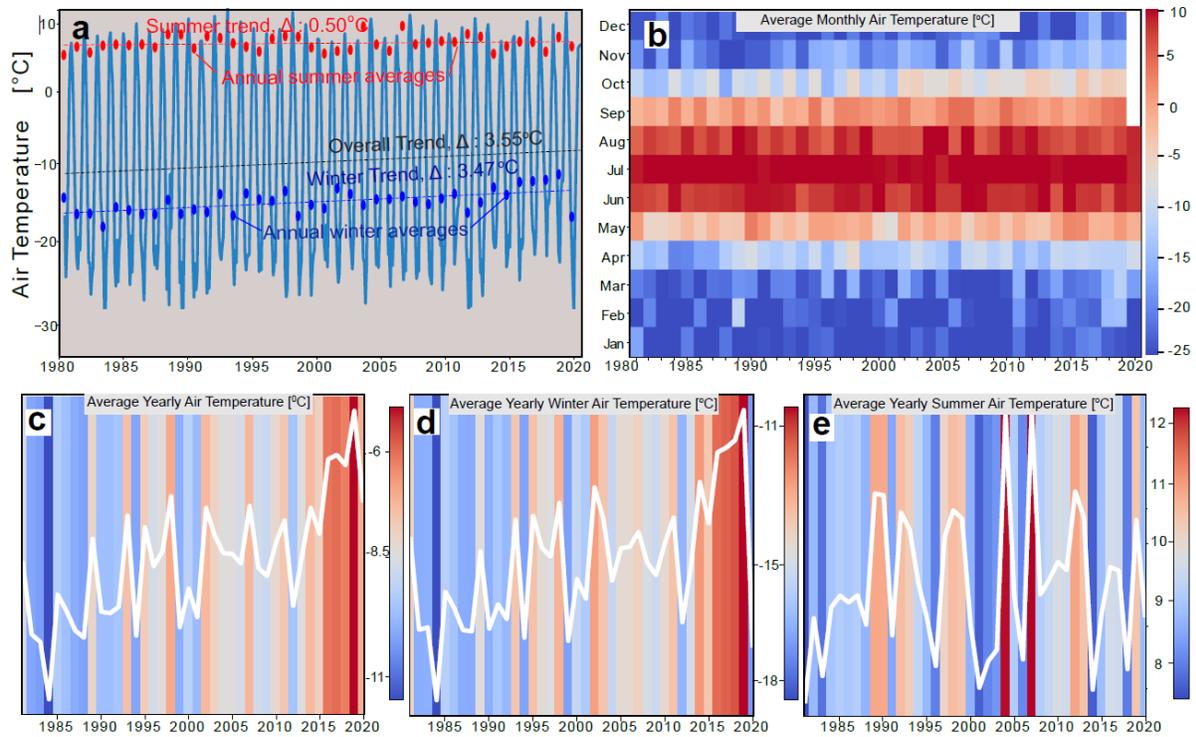

**Fig. S1.** Long term climate trends at the Colville River watershed, 1981-2020, from the NASA DAAC ABoVE Snowmodel dataset. **(a)** Annual mean, winter (September–May), and summer (June–August) air temperature trends with Theil–Sen trend estimates (Δ), **(b)** Heatmap showing monthly mean temperatures. Each block shows average temperature for a single month. **(c)** Heatmap showing yearly mean temperatures for each year, **(d)** Heatmap showing yearly mean winter temperatures and **(e)** Heatmap showing yearly mean summer temperatures.

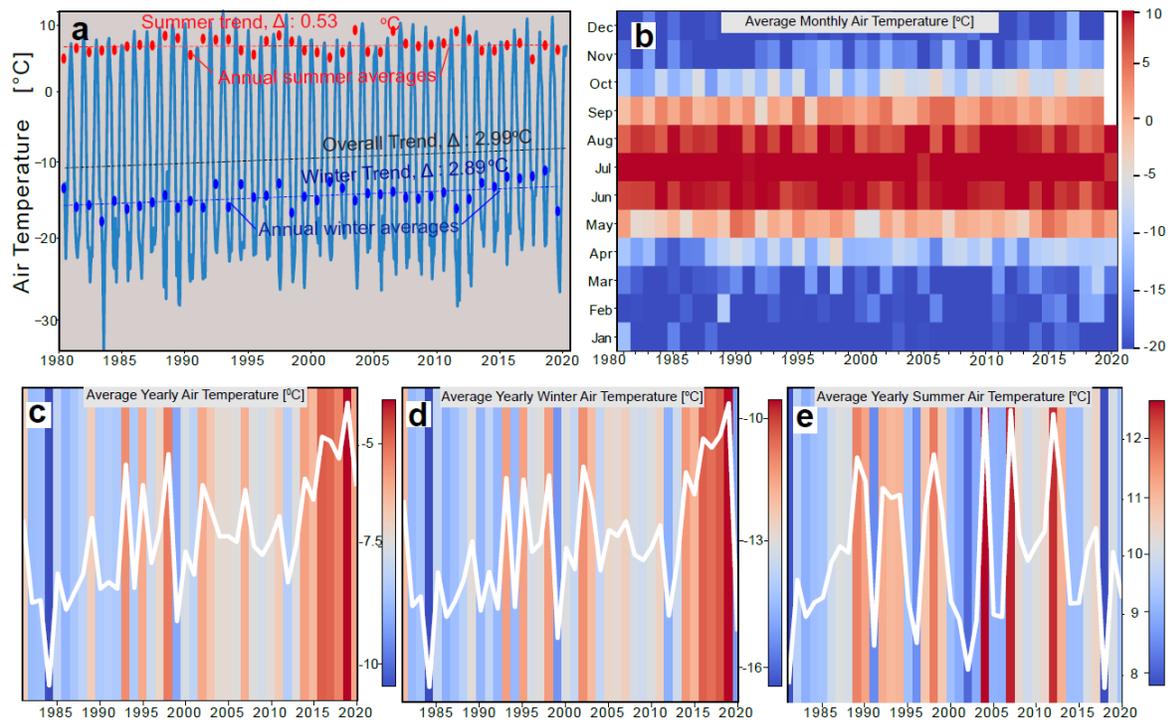

**Fig. S2**. Long term climate trends at the Sagavanirktok River watershed, 1981-2020, from the NASA DAAC ABoVE Snowmodel dataset. (a) Annual mean, winter (September - May), and summer (June - August) air temperature trends with Theil-Sen trend estimates (Δ), (b) Heatmap showing monthly mean temperatures. Each block shows average temperature for a single month. (c) Heatmap showing yearly mean temperatures for each year, (d) Heatmap showing yearly mean winter temperatures and (e) Heatmap showing yearly mean summer temperatures.

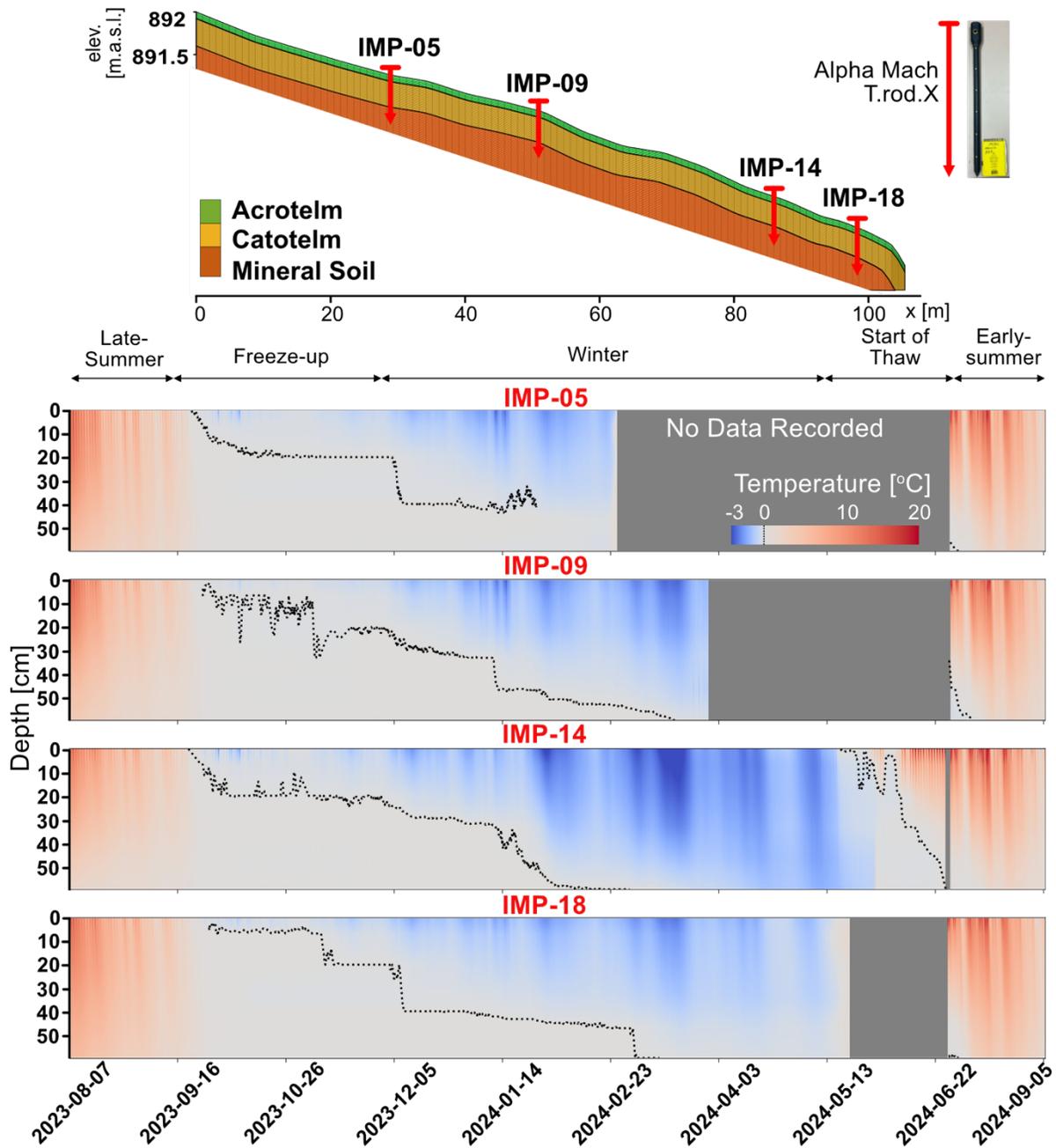

**Fig. S3.** Vertical soil temperature profiles from thermistor rod (T.rod.X®) arrays installed at four locations (IMP-05, IMP-09, IMP-14, and IMP-18) along the Imnavait Creek hillslope riparian transect. Points show measurements from 0-50 cm depth spanning late summer, freeze-up, winter, thaw onset, and early summer periods in 2023-2024. Elevation profile (top panel) indicates positions of the acrotelm, catotelm, and mineral soil layers at each site. Intervals colored gray indicate periods with no recorded data due to instrument malfunction during winter.

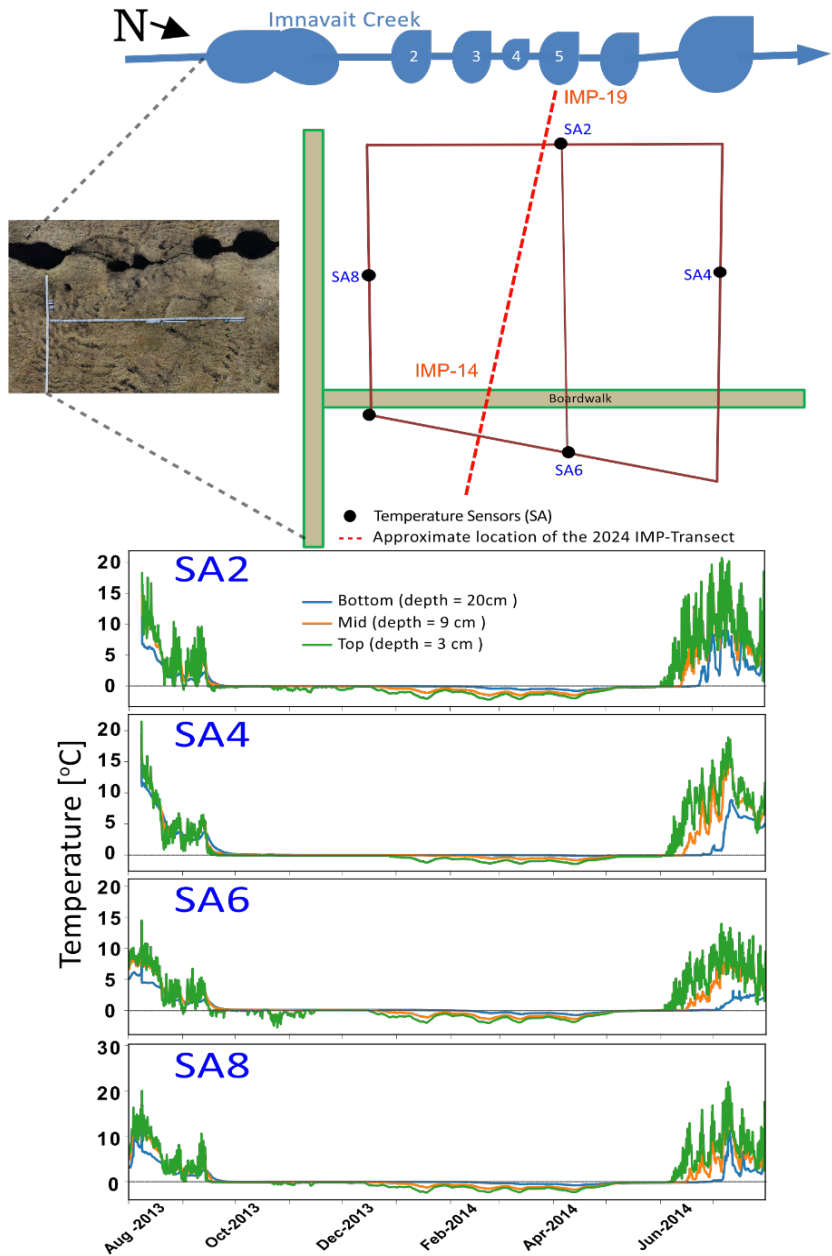

**Fig. S4.** Soil temperature observations from the 2013 - 2014 winter season highlighting the zero-curtain period (ZCP). Panels show temperature records from sensors SA2, SA4, SA6, and SA8 installed across the riparian zone of the Imnavait Creek hillslope, coinciding with the location of the instrumented transect (IMP), which we modeled. Persistent near-zero temperatures during winter indicate the presence of ZC zones in the soil during winter. ZCP for SA2, SA4, SA6 and SA8 are 190, 288, 183, 186 days, respectively. These values overlap with the modeled ZCP duration in the riparian zone for the 2013-2014 winter These values overlap with the modeled ZCP duration in the riparian zone for the 2013-2014 winter season.

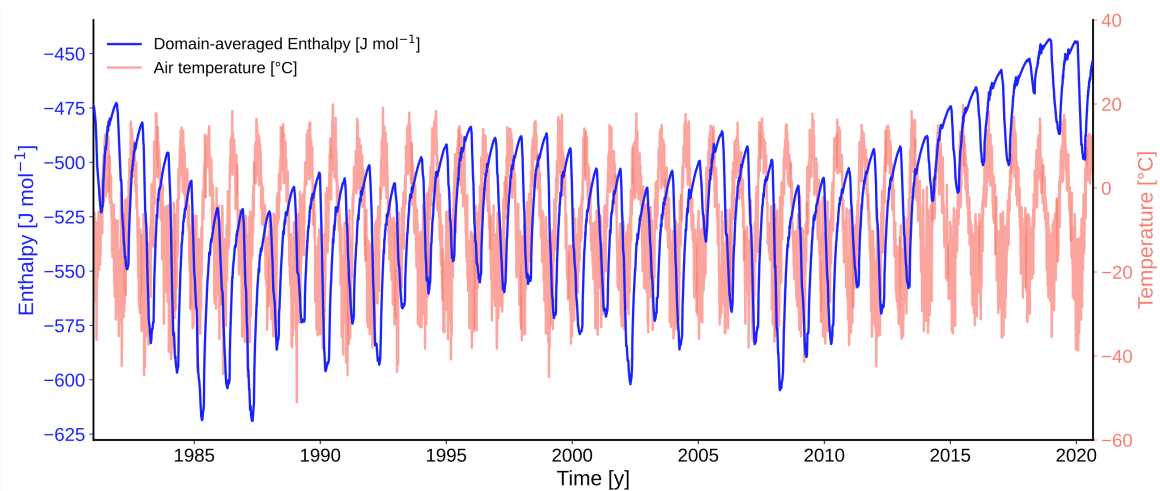

**Fig. S5.** Daily domain-averaged liquid water enthalpy [J mol$^{-1}$] and air temperature [°C] from 1981-2020. Local minima correspond to post-winter minimum energy states, while local maxima represent pre-summer heat content. The interannual variability reflects that domain averaged enthalpy is providing long-term energy balance memory, as opposed to instantaneous signals, i.e., daily air temperature.

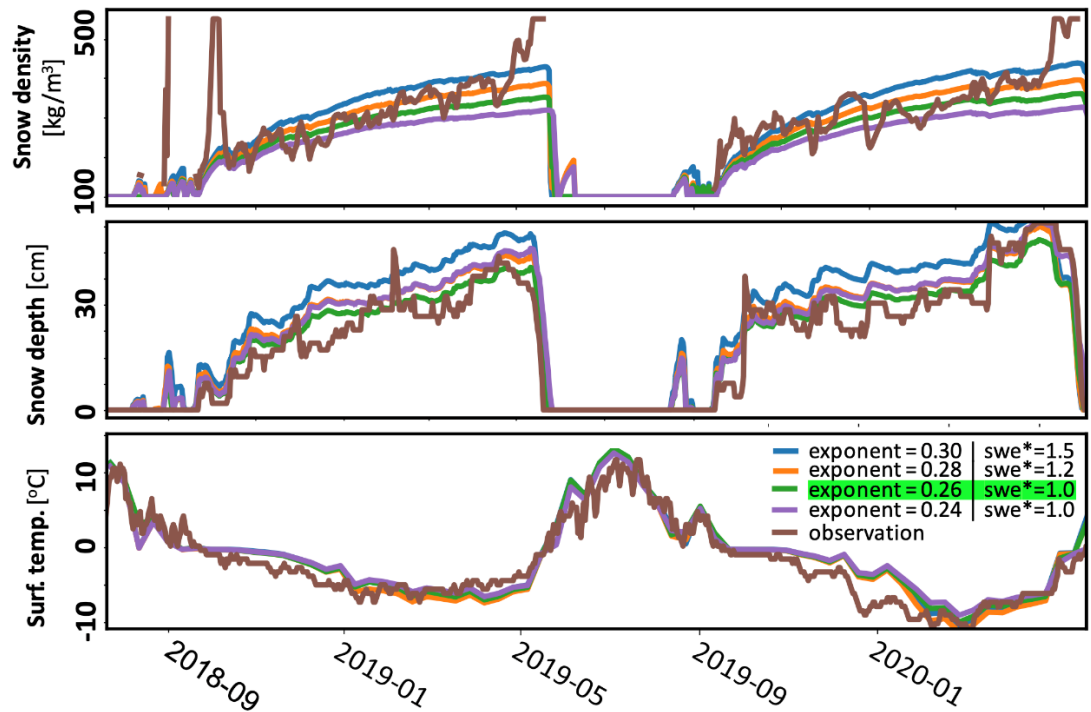

**Fig. S6.** Calibration of snowpack thermal transport in the model against observations from the nearby SNOTEL station (2011-2020). Comparison of modeled and observed snow density, snow depth, and surface temperature for different snow compaction exponents and winter snow water equivalent (SWE) scaling factors. The best-fit parameterization (exponent = 0.26, SWE* = 1.0) most closely matches observed seasonal evolution of snow properties.

**Mov. S1.** Movie showing hydrological conditions over a zero-curtain period from August 31, 2017 to May 26, 2018. Top panels show the liquid (blue), gas (black) and ice (gray) saturation and temperature (red) profiles till 1m depth at three locations x = 2 m (upslope), 60 m (midslope) and 105 m (downslope) along the transect. Lower panels show liquid water saturation (color scale), snow or ponded surface extent (white or blue overlay). The panel below that shows the groundwater flux directions (arrows), and magnitude (color scale in $\log_{10}$[cm/d]). The lowest panel shows the temperature [°C] for the subsurface (color scale) – note that cells with 0 °C are marked black in the color scale.

**Mov. S2.** Movie showing summer hydrological activity from May 1 to October 31, 2019. Top panel shows timeline of hydrological forcings like rain and snow precipitation on the top axis, and transect outflow on the bottom axis. The panel below that show the liquid (blue), gas (gray), and ice (gray) saturation and groundwater flux (brown) profiles till 1 m depth at three locations x = 5 m (upslope), 60 m (midslope) and 105 m (downslope) along the transect. The panel below that shows liquid water saturation (color scale), snow or ponded surface extent (white or blue overlay). The panel below that shows the groundwater flux directions (arrows), and magnitude (color scale in $\log_{10}$[cm/d]). The lower panel shows the direction and magnitude of surface normal fluxes (blue) and surface tangential fluxes (orange).